\documentclass[journal=langd5, manuscript=article]{achemso}
\usepackage{makecell}
\usepackage{chemformula}
\usepackage{achemso}
\usepackage{amsmath}    
\usepackage{verbatim}   
\usepackage{color}      
\usepackage{subfig}  
\usepackage{hyperref}   
\usepackage{amssymb}
\usepackage{amsthm}
\usepackage{amsfonts}
\usepackage{mathrsfs}
\usepackage{float}
\usepackage[version=4]{mhchem}
\usepackage[separate-uncertainty=true]{siunitx}
\usepackage{cleveref}
\usepackage{soul}
\usepackage{setspace}
\usepackage{multirow}
\usepackage{chemfig}
\usepackage{graphicx, caption}
\usepackage{siunitx}
\usepackage{multirow}
\usepackage{color,colortbl}

\usepackage{xr}
\makeatletter
\newcommand*{\addFileDependency}[1]{
  \typeout{(#1)}
  \@addtofilelist{#1}
  \IfFileExists{#1}{}{\typeout{No file #1.}}
}
\makeatother

\definecolor{Gray}{gray}{0.9}
\definecolor{ImportantCalculation}{RGB}{130,120,183}
\definecolor{bluebell}{rgb}{0.64, 0.64, 0.82}
\definecolor{lavenderblue}{rgb}{0.8, 0.8, 1.0}
\DeclareCaptionFormat{lower}{#1#2\lowercase{#3}\par}
\usepackage{siunitx}

\usepackage[color=lavenderblue]{todonotes} 
\usepackage{booktabs}
\usepackage{lscape}

\graphicspath{{./}}
\title{Ultrasound Propagation in Water-Sorbing \\Carbon Xerogel}
\author{Ashoka Karunarathne}
\affiliation{Otto H. York Department of Chemical and Materials Engineering, \\ New Jersey Institute of Technology, University Heights, Newark, NJ 07102, USA}

\author{Stephan Braxmeier}
\affiliation{Center for Applied Energy Research e.V. (CAE), Magdalene-Schoch-Strasse 3, 97074 Wuerzburg, Germany}

\author{Boris Gurevich}
\affiliation{Center for Exploration Geophysics, School of Earth and Planetary Sciences, Curtin University, 26 Dick Perry Avenue,
Kensington, Western Australia 6151, Australia} 

\author{\\Alexei F. Khalizov}
\affiliation{Department of Chemistry and Environmental Science,
New Jersey Institute of Technology,
323 Dr. Martin Luther King Jr. Blvd, Newark, NJ 07102, USA}
\alsoaffiliation{Otto H. York Department of Chemical and Materials Engineering, \\ New Jersey Institute of Technology, University Heights, Newark, NJ 07102, USA}

\author{Gudrun Reichenauer}
\affiliation{Center for Applied Energy Research e.V. (CAE), Magdalene-Schoch-Strasse 3, 97074 Wuerzburg, Germany}
\author{Gennady Y. Gor}
\affiliation{Otto H. York Department of Chemical and Materials Engineering, \\ New Jersey Institute of Technology, University Heights, Newark, NJ 07102, USA}
\email{gor@njit.edu}

\date{\today}
\usepackage{setspace}
 
\begin{document}
\setstretch{1.75}

\maketitle

\newpage
\begin{abstract}
Adsorption of water vapor in nanoporous carbons is rather complex due to an interplay between the pore structure and surface chemistry of these materials. Deciphering the mechanism of adsorption requires the knowledge of the spatial distribution and the filling fraction of the adsorbed water. Characteristics of ultrasonic wave propagation through a nanoporous sample provides a wealth of information such as properties of fluids confined in the pores and spatial distribution of adsorbate in the pores. Here we studied water vapor adsorption on carbon xerogel, with a bimodal pore size distribution consisting of micropores (1 nm) and mesopores (8 nm) and overall porosity 68\%. The relative humidity was increased in steps and the amount of water adsorbed by the sample was measured gravimetrically at each step, producing a type V isotherm, characteristic of water adsorption to weakly interacting carbon nanopores. Concurrently, we recorded the waveforms for longitudinal and transverse ultrasonic waves that passed through the sample at each saturation step. These measurements provided the wave amplitudes and speeds as a function of relative humidity, and these data were used to calculate the elastic moduli and ultrasonic attenuation of the xerogel-water composite. Analysis of the elastic moduli evolution suggested that confined water shows nearly bulk-like properties in mesopores, while in micropores its modulus noticeably differs from that of bulk water, consistent with theoretical predictions. Furthermore, the observed increase in ultrasonic attenuation during micropore filling indicated the spatial heterogeneity of the water-filled pore space within the overall sample volume. Overall, this study demonstrates the successful utilization of nondestructive ultrasonic testing to probe the fluid adsorption mechanism in a nanoporous medium and the properties of adsorbed phase.

\end{abstract} 
\newpage

\section{Introduction}

Acoustic studies of vapor-sorbing porous materials go back at least four decades, to the experiments by Murphy, who investigated the attenuation of low-frequency acoustic waves (300~Hz to 14~kHz) as a function of water saturation for Massilon sandstone and Vycor glass samples~\cite{Murphy1982}. Subsequent work focused on various aspects of the complex behavior exhibited by the fluid-porous medium composite. For instance, Warner and Beamish~\cite{Warner1988} studied ultrasound propagation in a Vycor glass sample during nitrogen adsorption at 77 K, proposing ultrasound as an alternative tool for measuring adsorption isotherm. Page et al.~\cite{Page1993, Page1995} used Vycor glass to measure the propagation speed and attenuation of an ultrasound wave during adsorption and desorption of n-hexane, to explore pore-filling and pore-emptying mechanisms and relate them to optical measurements on the same Vycor material. More recently Schappert and Pelster published a series of works, reporting ultrasound propagation in Vycor glass in the course of adsorption/desorption of various vapors at ambient and cryogenic temperatures~\cite{Schappert2013JoP, Schappert2014, Schappert2015EPL, Schappert2018, Schappert2018liquid, Schappert2022, Schappert2024}. Combined adsorption-acoustic experiments were also used to study the change in mechanical properties caused by water sorption in real rock samples, including sandstone, limestone, shale, and granite, along with model glass-bead based porous medium~\cite{Pimienta2014, Ilin2020, Khajehpour2020, Tiennot2020, Mikhaltsevitch2021, Ogunsami2021, Gao2022, Gao2023, Wu2023}. Lastly, the authors of this paper obtained a wealth of information by combining water sorption with ultrasound propagation for a diverse set of samples represented by sandstone~\cite{Yurikov2018}, Vycor glass~\cite{Ogbebor2023}, and macroporous polymer composites~\cite{Karunarathne2024}. This body of work has brought attention to the behavior of acoustic waves during vapor sorption in \textit{nanoporous} materials, since fluids in nanopores show many unique properties~\cite{Gor2014, Maximov2018, Dobrzanski2021}, some of which can be retrieved from the ultrasound-based measurements. Vycor glass is the only nanoporous material that has been studied by combined adsorption–ultrasonics in great detail. This is due to its availability in the form of monolithic samples, its accessible and well-defined pore space (7-8 nm diameter), and its hydrophilic pore wall surface. 

Similar to Vycor, nanoporous carbons show a complex adsorption behavior that has been widely studied to understand the mechanism and governing factors in order to utilize these carbons in materials applications~\cite{Liu2017review}, such as adsorption and separation, energy storage, catalysis, hydrogen storage and environmental remediation~\cite{Benzigar2018,Memetova2022,Kostoglou2022}. The adsorption mechanism and the characteristics of isotherms depend on the properties of the vapor being adsorbed and also vary between the materials with different pore structures and surface chemistries, which are made available by modern synthesis methods. Indeed, nowadays nanoporous carbon materials with unique physical and chemical properties can be prepared, including well-controlled porous structure, high specific surface areas, electronic and ionic conductivity, and enhanced mass transport and diffusion. 

For water vapor, the sorption isotherms at low to intermediate humidities are S-shaped with a hysteresis loop following type V classification, corresponding to relatively weak adsorbent–adsorbate interactions~\cite{Thommes2015}. To unveil the detailed mechanisms of water adsorption on carbons, additional experimental techniques are widely utilized. For instance, Iiyama et al.~\cite{Iiyama1995} and Kaneko et al.~\cite{Kaneko1999}, based on X-ray diffraction (XRD) measurements and electron radial distribution function analysis, reported formation of an ordered assembly structure of water adsorbed in carbon micropores. Bahadur et al.~\cite{Bahadur2017} studied the kinetics of water adsorption in ultramicroporous carbon using in-situ small angle neutron scattering (SANS), confirming that the adsorption of water occurs via cluster formation and showing the variations of geometry and the concentration of the clusters over the adsorption process. Nuclear magnetic resonance (NMR) studies conducted by Wang et al.~\cite{Wang2014NMR} proposed a two-step filling mechanism from the measured variation of spin-lattice relaxation time. To our knowledge, there have not been attempts to complement water adsorption experiments in nanoporous carbons with ultrasound propagation measurements.

Here we used the combined ultrasonic pulse transmission--vapor adsorption technique to investigate the characteristics of ultrasound wave propagation through water-sorbing carbon xerogel, a synthetic nanoporous carbon with tunable material properties, available as a monolithic sample. The xerogel in this study had a bimodal pore size distribution, comprised of micropores and mesopores, giving rise to two-step pore filling during water adsorption, as shown previously~\cite{Thommes2013}. The unique characteristics of ultrasonic wave propagation allowed us to shed light on the mechanism of water adsorption, associated adsorption-induced sample deformation and the bulk modulus of the water confined in nanopores.

\section{Results}
\subsection{Xerogel material properties}
\begin{figure}[H]
\centering
\subfloat[]{\includegraphics[width=0.45\textwidth]{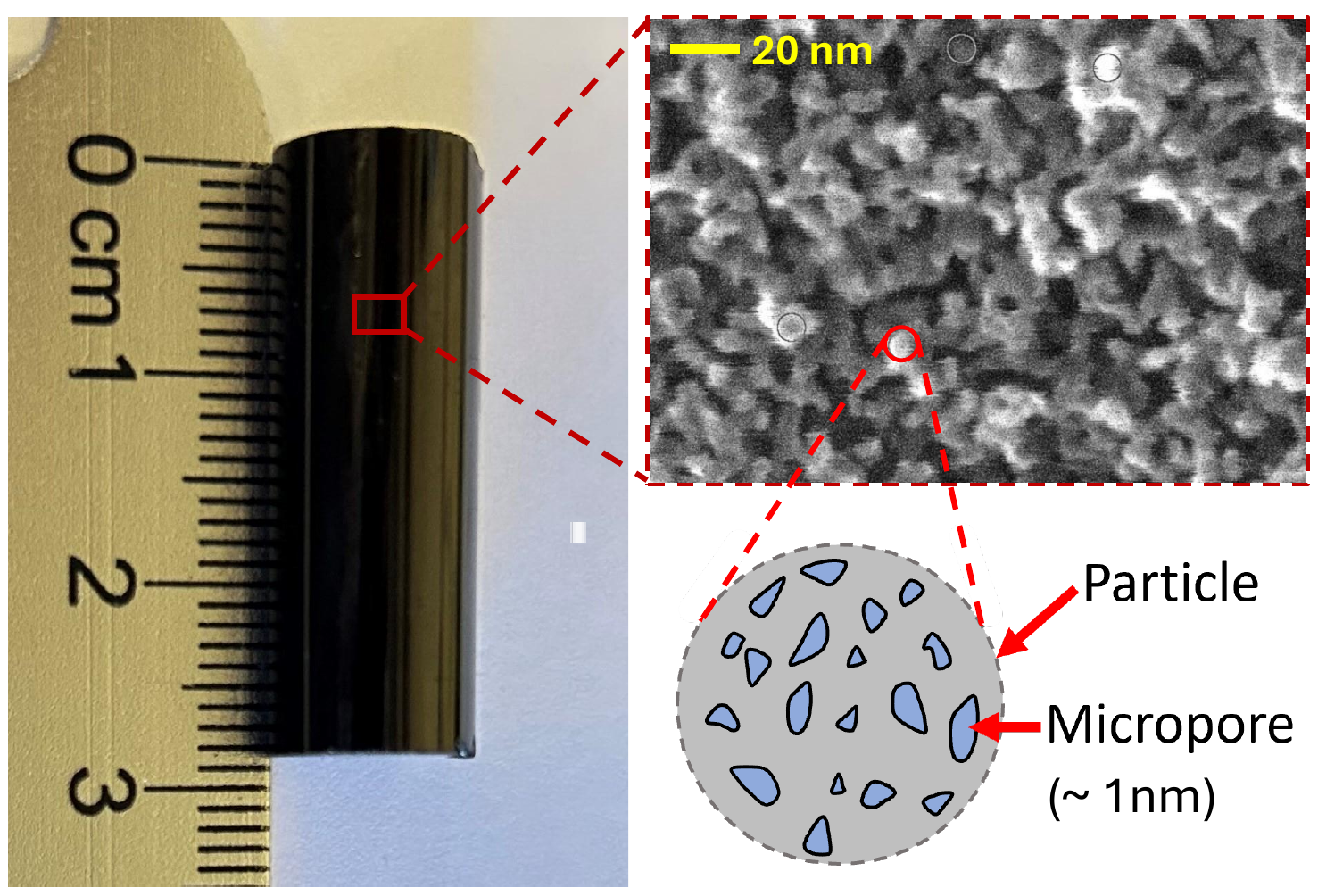} \label{fig:SEM-sample}}
\subfloat[]{\includegraphics[width=0.45\textwidth]{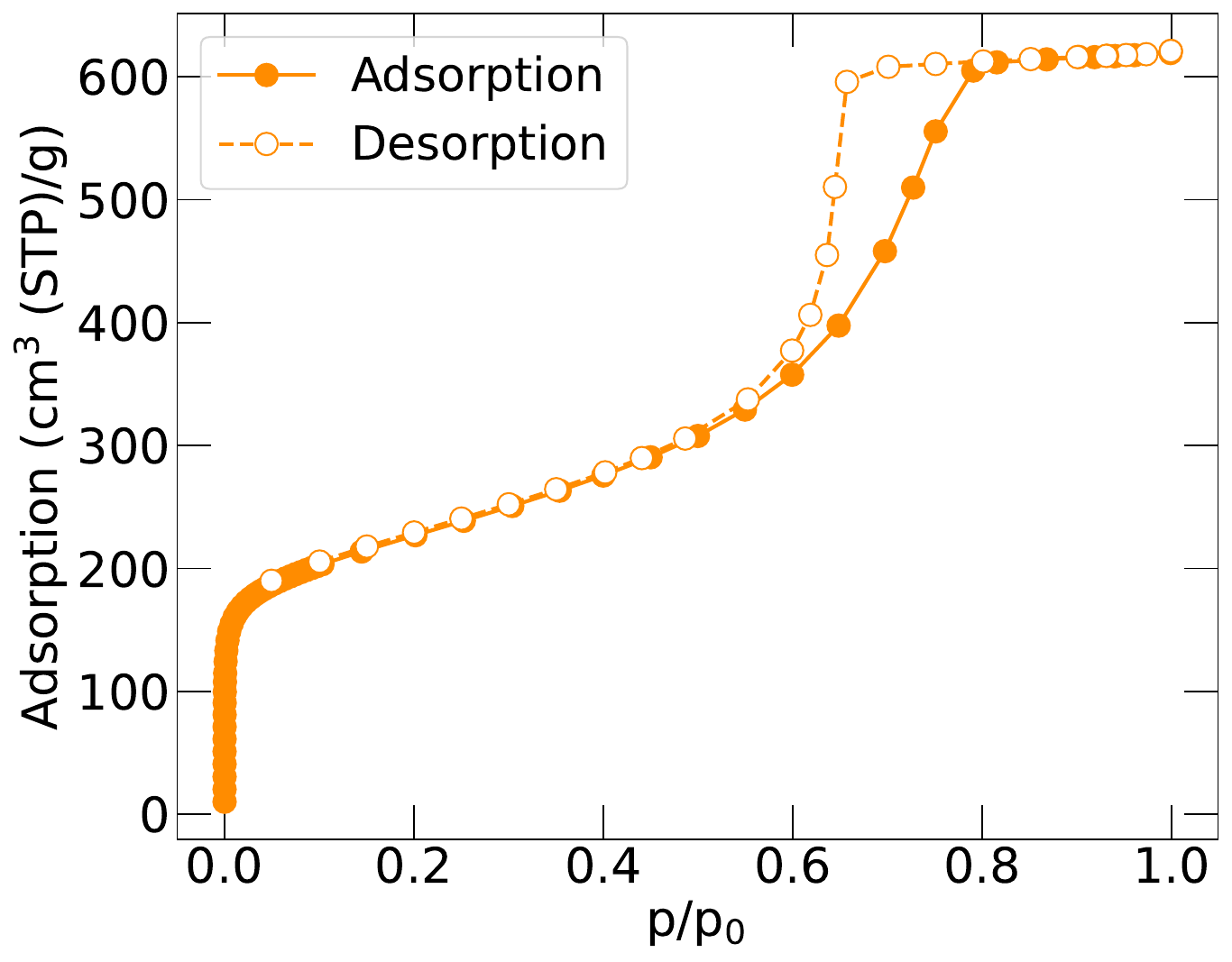} \label{fig:isotherm-N2}}

\subfloat[]{\includegraphics[width=0.45\textwidth]{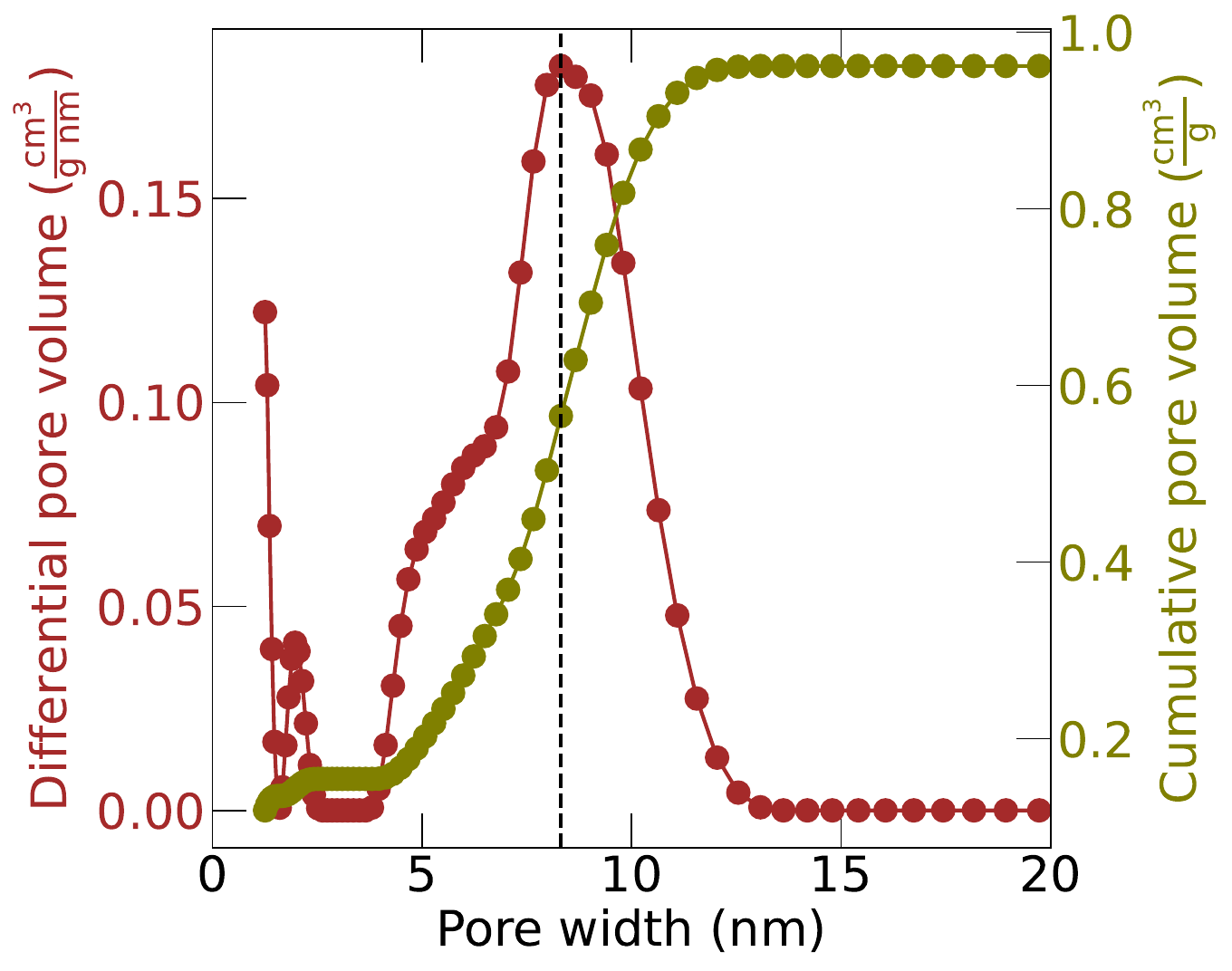} \label{fig:PSD-mesopore}}
\subfloat[]{\includegraphics[width=0.45\textwidth]{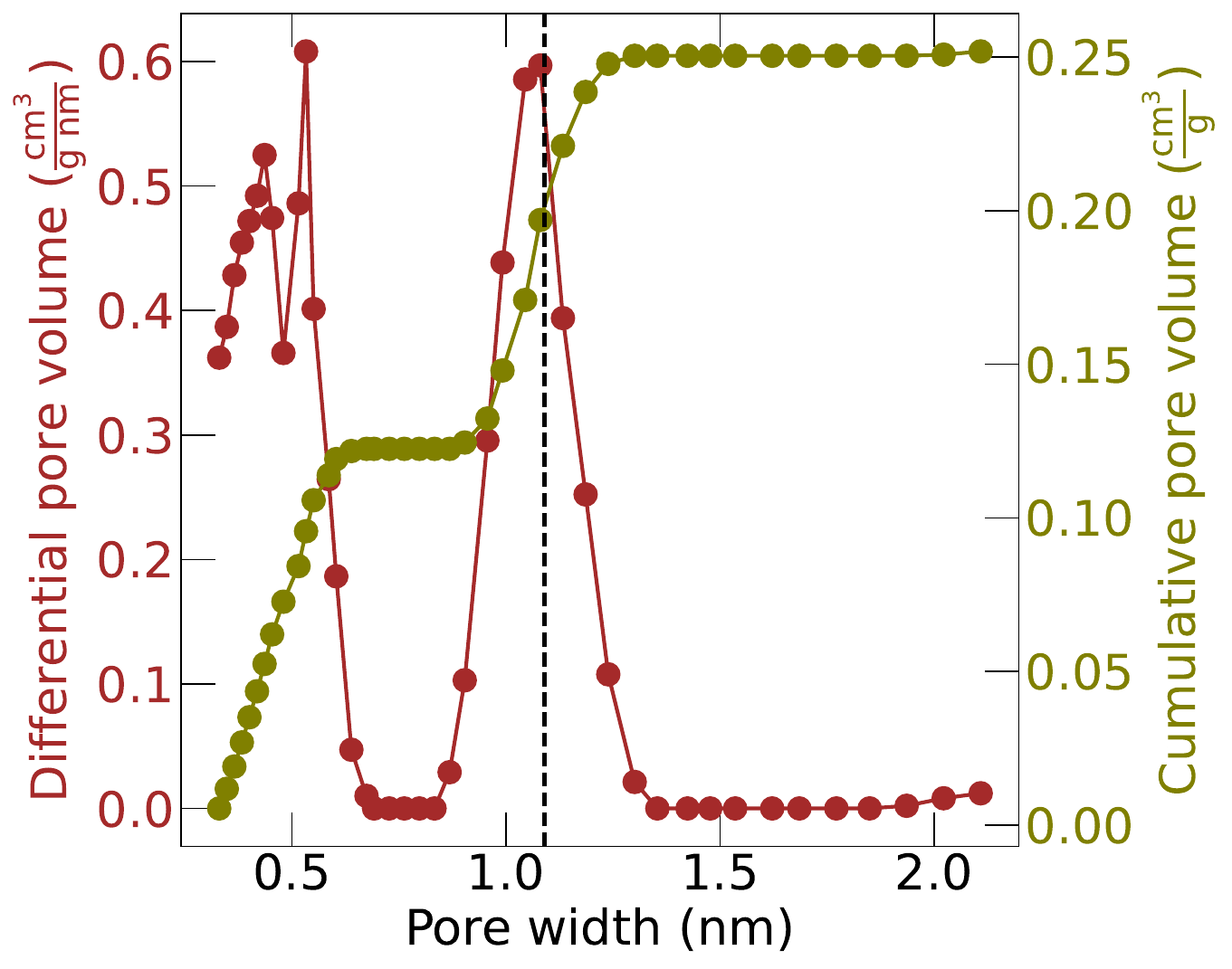} \label{fig:PSD-micropore}}

\caption{\textbf{Xerogel pore structure}. (a) Monolithic synthetic carbon sample used in this study, its SEM image, and a schematic depiction of its microstructure, (b) nitrogen sorption isotherm at 77 K, (c) differential and cumulative mesopore size distributions obtained from a non local DFT (NLDFT) Kernel for nitrogen on carbon with cylindrical pore geometry, and (d) differential and cumulative micropore size distributions obtained using a combination of two NLDFT models from \ce{CO2} and \ce{N2} sorption measurements (\ce{CO2} at 273 K in carbon and \ce{N2} at 77 K on carbon slit pores) using the “NLDFT advanced PSD” tool of Microactive by Micromeritics, GA, USA.}
\label{fig:poresizes}
\end{figure}

Fig.~\ref{fig:SEM-sample} shows a SEM image and a schematic depiction of the carbon xerogel microstructure, which consists of a network of interconnected particles with average size, $d_{\rm part}$ of $8 ~\rm nm$. The cavities in the pore network between the particles fall within the range of the mesopores ($\sim$8 nm) but there are also intraparticle micropores ($\sim$1 nm). These micropores are formed during sample carbonization, where the organic polymer precursor undergoes thermal decomposition, leading to the formation of small, interconnected voids inside the resulting carbon particles. Presence of both micropores and mesopores in the sample can be  seen in the \ce{N2} sorption isotherm (Fig.~\ref{fig:isotherm-N2}), which is clearly divided into two regions. The micropore region corresponds to a steep rise at low relative pressure $p/p_{0}$ just above zero and capillary condensation in mesopores corresponds to a steep rise at $p/p_{0}$ around 0.6. Figs.~\ref{fig:PSD-mesopore} and \ref{fig:PSD-micropore} show calculated pore size distributions for mesopores and micropores, respectively, and the material properties of the xerogel used in this study are listed in Table~\ref{Table:materials properties}.  

\captionsetup[figure]{position=bottom,skip=12pt}
\captionsetup[subfigure]{
  position=top,
  captionskip=1pt,
  singlelinecheck=false,
}

\begin{table}[]
\begin{tabular}{ccccccccc}
\hline

\makecell{$\rho_{\rm b}$\\(g/cm$^{3}$)} &\makecell{$\phi$\\(\%)} & \makecell{$S_{\rm BET}$\\(m$^{2}$/g)}& \makecell{$V_{\rm pores}$\\(cm$^{3}$/g)}&\makecell{$V_{\rm micropore}$\\ (cm$^{3}$/g)}&\makecell{$\phi_{\rm micropores}$\\(\%)}&\makecell{ $d_{\rm part}$\\ (nm)}&\makecell{$d_{\rm meso}$\\(nm)}&\makecell{$d_{\rm mic}$\\(nm)}\\
\hline
0.707&68&800.3&0.959&0.124&8.8&8.1&8.3&1.09\\
\hline
\end{tabular}
\caption{Materials properties of the carbon xerogel samples used in this study. ($\rho_{\rm b}$: bulk density, $\phi$: porosity, $S_{\rm BET}$: specific BET surface area, $V_{\rm pores}$: total specific pore volume, $V_{\rm micropore}$: specific micropore volume, $\phi_{\rm micropores}$: micropore porosity, $d_{\rm part}$: average particle size, $d_{\rm meso}$: average mesopore size, and $d_{\rm mic}$: average micropore size) }
\label{Table:materials properties}
\end{table} 

\subsection{Equilibration of xerogel sample at different vapor saturations}
\begin{figure}[H]
\centering

\subfloat[]{\includegraphics[width=0.35\textwidth]{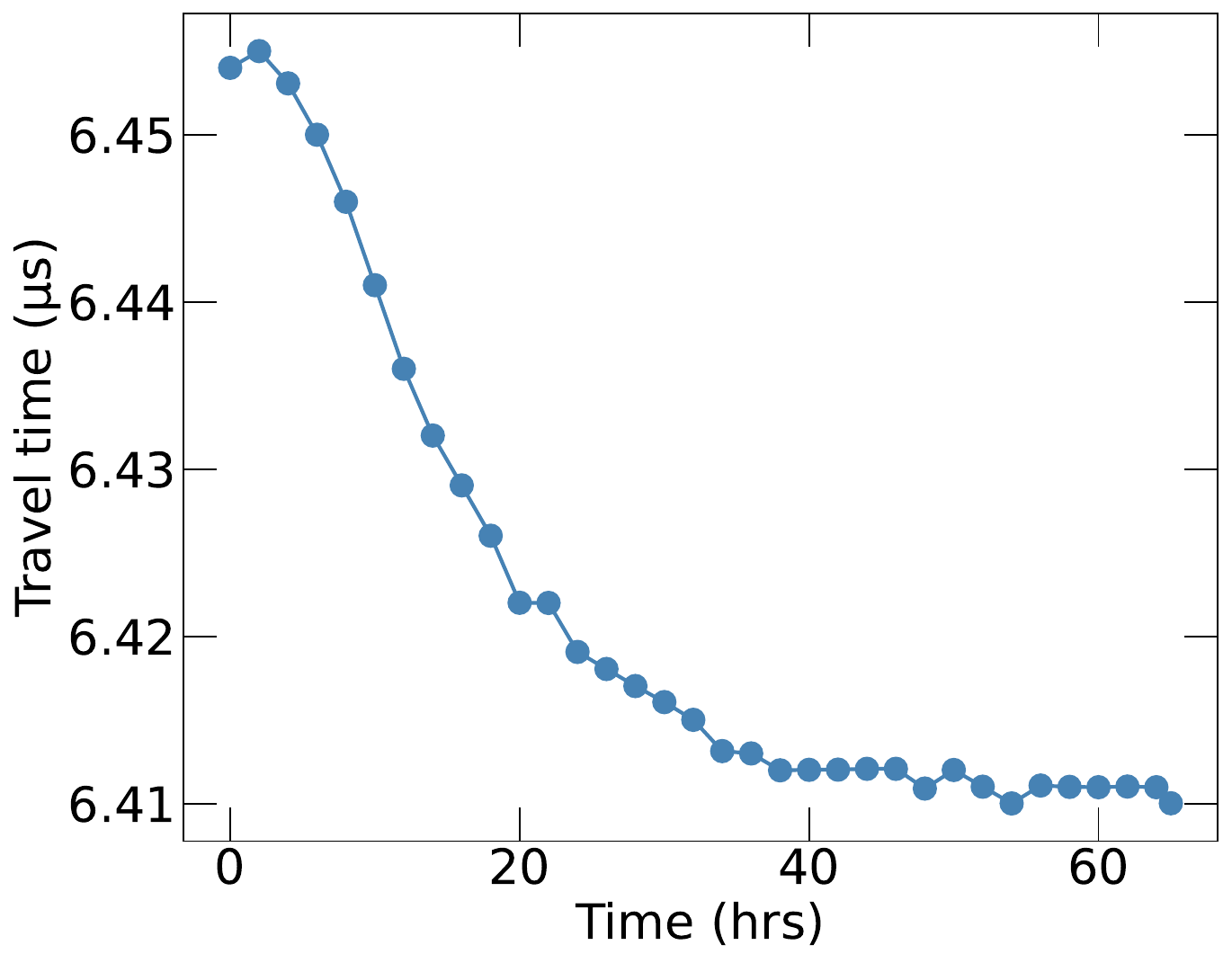} \label{Fig:equilibration-RH 12} }
\subfloat[]{\includegraphics[width=0.35\textwidth]{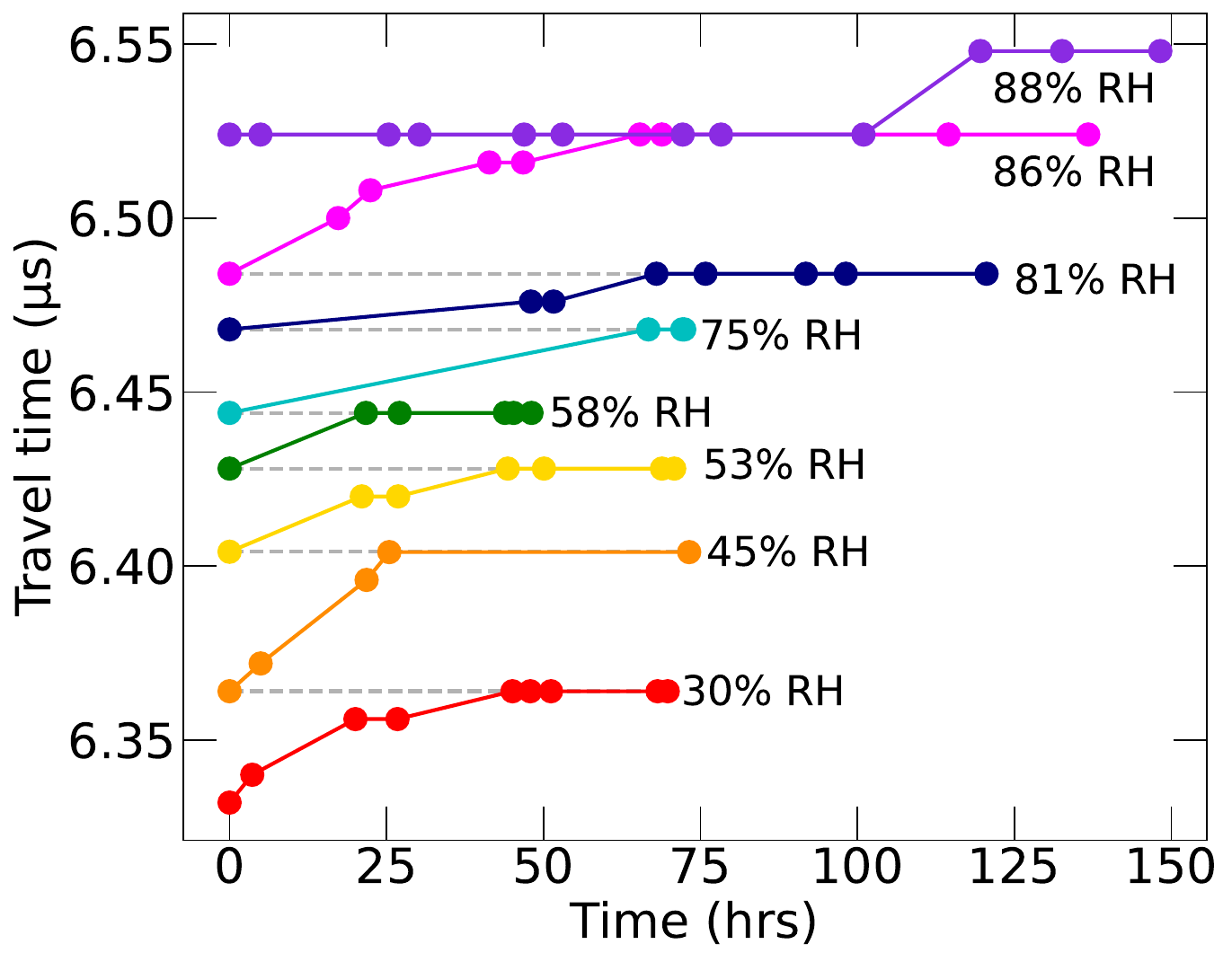}\label{fig:equilibration-RH all}}

\subfloat[]{\includegraphics[width=0.35\textwidth]{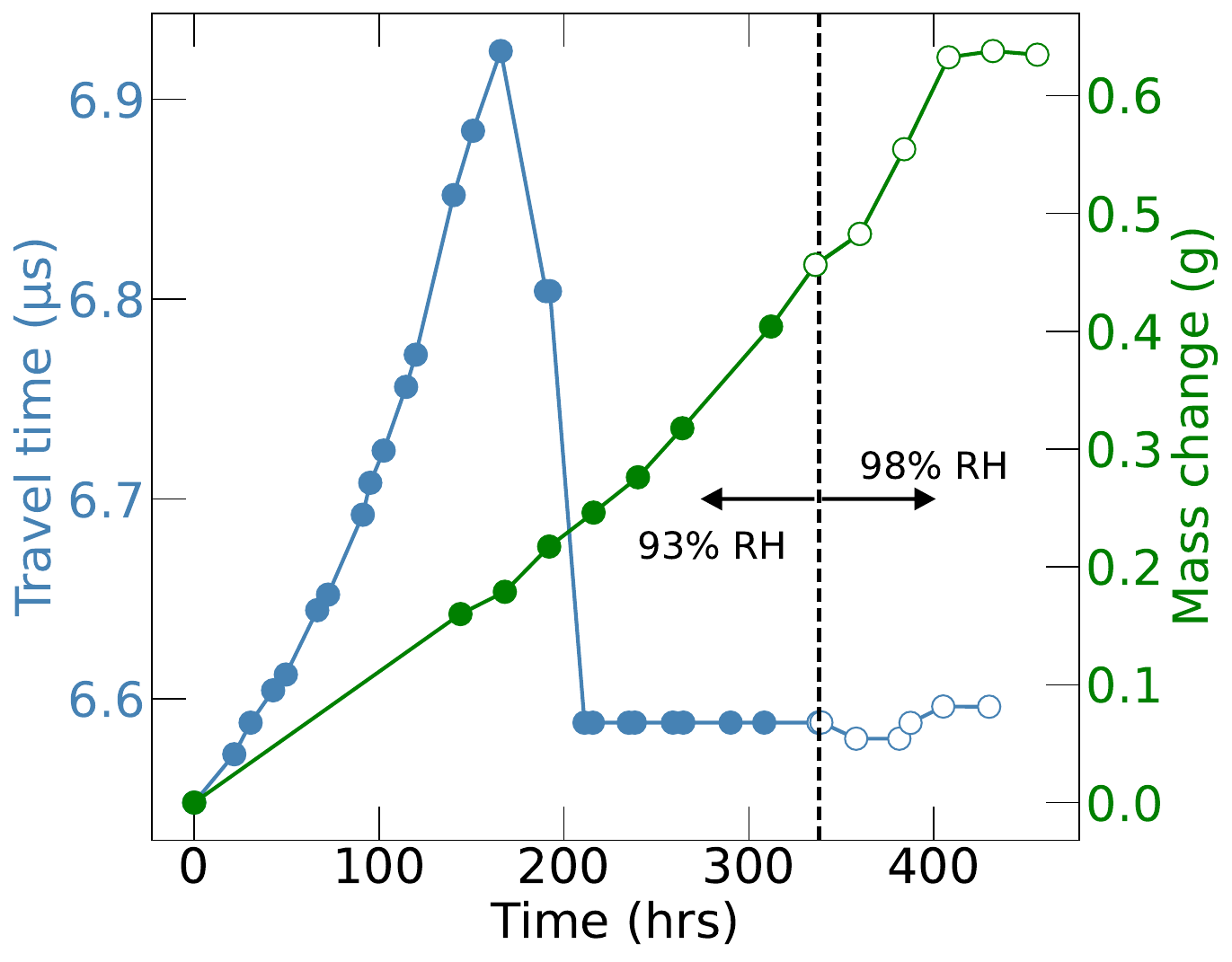}\label{fig:RH93&98-time&mass}}
\subfloat[]{\includegraphics[width=0.35\textwidth]{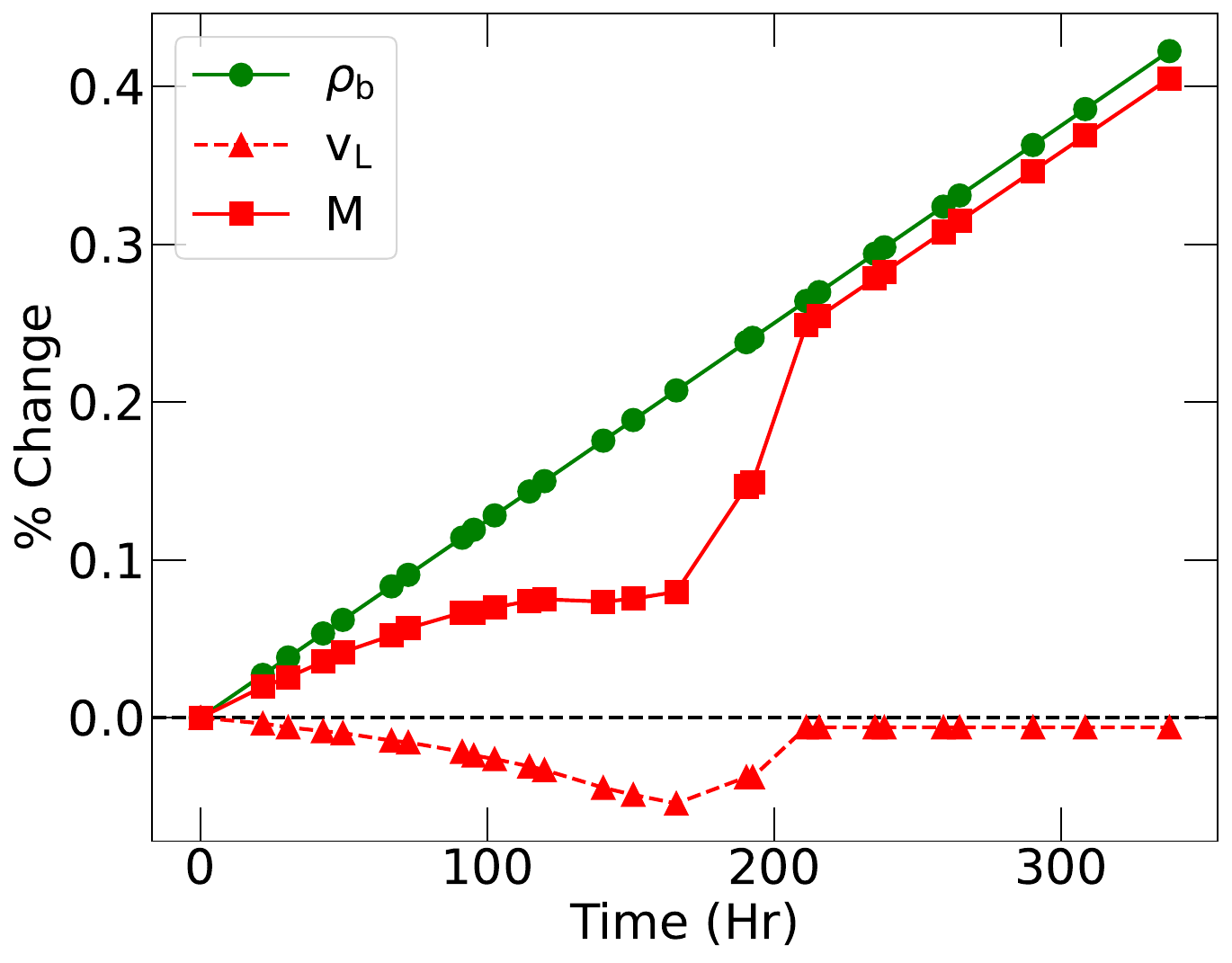}\label{fig:RH93-VMm}}

\subfloat[]{
\includegraphics[width=0.35\textwidth]{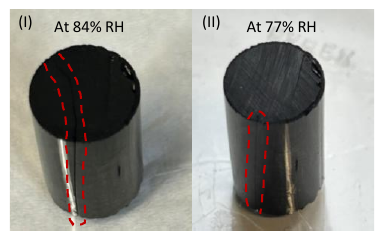} \label{fig:sample-crack}}
\subfloat[]{
\includegraphics[width=0.35\textwidth]{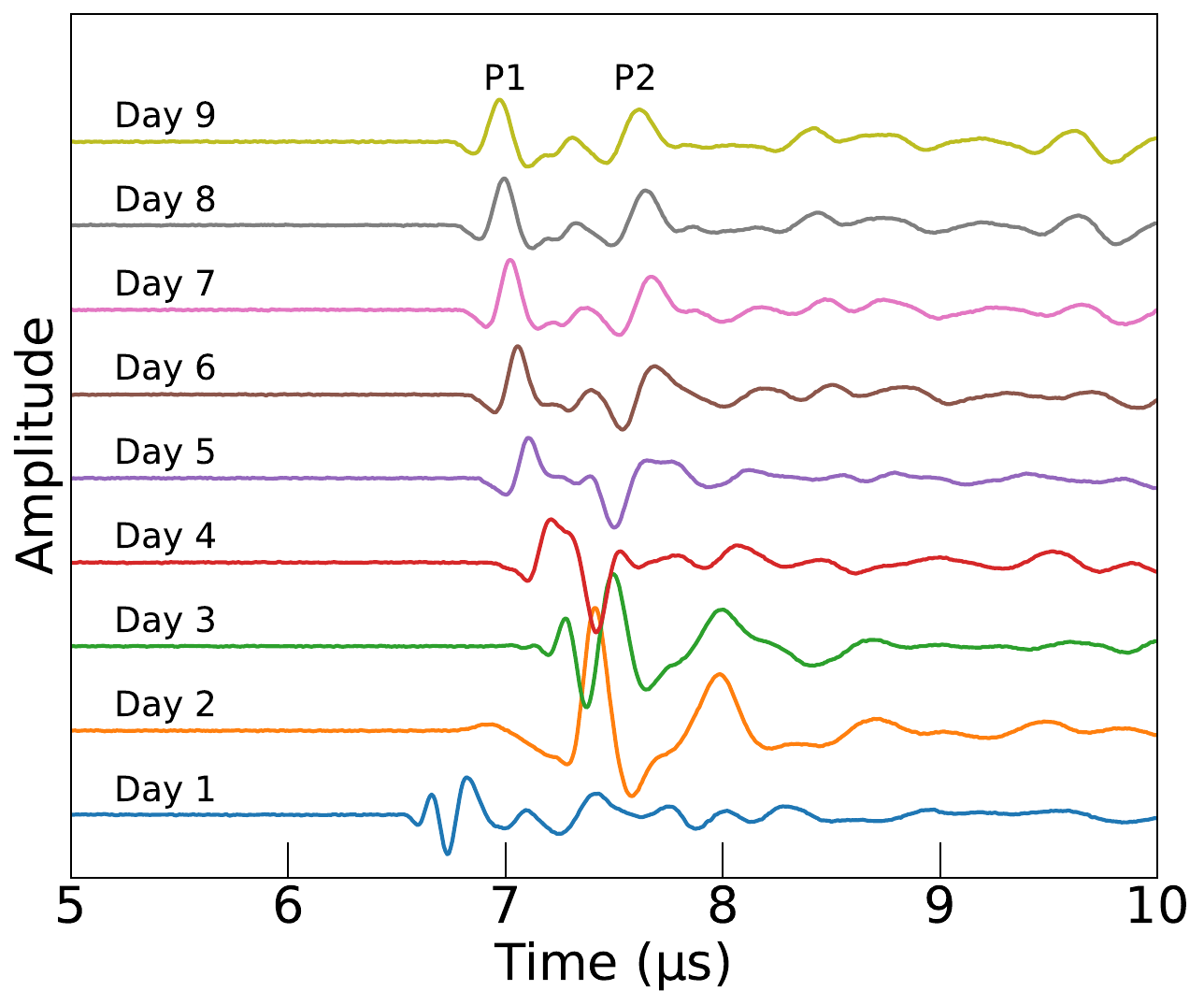} \label{fig:waveform-crack}}
    \caption{\textbf{Equilibration of water vapor saturation}. Variation of the travel time of the longitudinal wave with the exposure time of the sample (a) after changing RH from 48\% to 12\% and (b) to each RH level from 30\% to 88\%. (c) Variation of the travel time of the longitudinal wave and the sample mass with the exposure time of the sample to 93\% RH (filled markers) and 98\% RH (empty markers). (d) \%Change of longitudinal speed ($v_{\rm L}$), bulk density ($\rho_{\rm b}$) and the longitudinal modulus ($M$) with the exposure time of the sample to 93\% RH. Note: the base time, $t=0$ is taken when the sample is exposed to 93\% RH after equilibration at 88\% RH. (e) Sample cracks observed at (I) 84\% RH and (II) 77\% RH during the desorption route. (f) Growth of additional longitudinal peak in the waveforms measured over the time at 84\% RH.}
    \label{fig:equilibration-RH}

\end{figure}

Due to the high specific surface area and large sample pore volume, the equilibration between the sample and its water vapor environment was slow. To determine the equilibration time, a combination of ultrasonic and gravimetric measurements was performed. Thus, while we are primarily interested in equilibrium properties, our measurements allowed us to evaluate adsorption kinetics as well. Fig.~\ref{Fig:equilibration-RH 12} shows the variation of the measured travel time of the longitudinal wave with the exposure time for the sample that started at ambient relative humidity (RH) of 48\% and then was exposed to 12\% RH. Based on the variation of the travel time, the estimated equilibration time at 12\% RH was nearly 60 hrs. After the equilibration, the sample was exposed to progressively higher RH in steps. Fig.~\ref{fig:equilibration-RH all} shows the variation of the travel time with the exposure time at each RH level, ranging from 30\% to 88\%. The starting point of each of curve is the end point obtained at the previous RH. Notably, when RH was raised to 93\%, the evolution in travel time does not show a monotonic behavior. Instead, as illustrated in Fig.~\ref{fig:RH93&98-time&mass}, there is an increase in travel time during the first $\sim$150 hrs of water adsorption, followed by a relatively sharp drop in travel time nearly to its initial level, and then the travel time remains constant after 200 hrs. Contrary to the ultrasound measurement, the mass measurement does not indicate an equilibration of the sample with water vapor, and even after 15 days of exposure at 93\% RH the sample mass continues to increase (Fig.~\ref{fig:RH93&98-time&mass}). Figure~\ref{fig:RH93-VMm} shows the  corresponding percent change in the longitudinal wave speed, along with the estimated sample density and longitudinal modulus, as a function of time after changing the RH from 88\% to 93\%. In agreement with the trends in travel speed and mass gain, the relative change in the ultrasound speed is much smaller than the change in the sample density. After 360 hours (15 days), the target humidity was adjusted to 98\% RH, and after additional $\sim$ 100 hrs of exposure, both travel time and mass reached a plateau (Fig.~\ref{fig:RH93&98-time&mass}). At this point, the adsorption experiment was deemed completed and the desorption route was initiated. 

During desorption, the sample cracking was visually noted at 84\% RH, and during further desorption, partial healing of the sample crack was observed at 77\% RH (Fig.~\ref{fig:sample-crack}). This sample cracking was also detected by the ultrasound measurement from the growth of an additional longitudinal peak in the waveforms obtained at 84\% RH as shown in Fig.~\ref{fig:waveform-crack} (day 2). The origin of the additional peak may be due to creating an additional path for wave propagation or reflection of waveform at the crack surface of the sample. The shear peak disappeared after the sample cracking. The observed loss of shear wave amplitude can be due to the discontinuity of the ultrasonic pathway for the shear waves at the crack interface. However, the measurements were continued with the cracked sample.

\subsection{Water sorption isotherms}
\begin{figure}[H]
\centering
\subfloat[]{\includegraphics[width=0.5\textwidth]{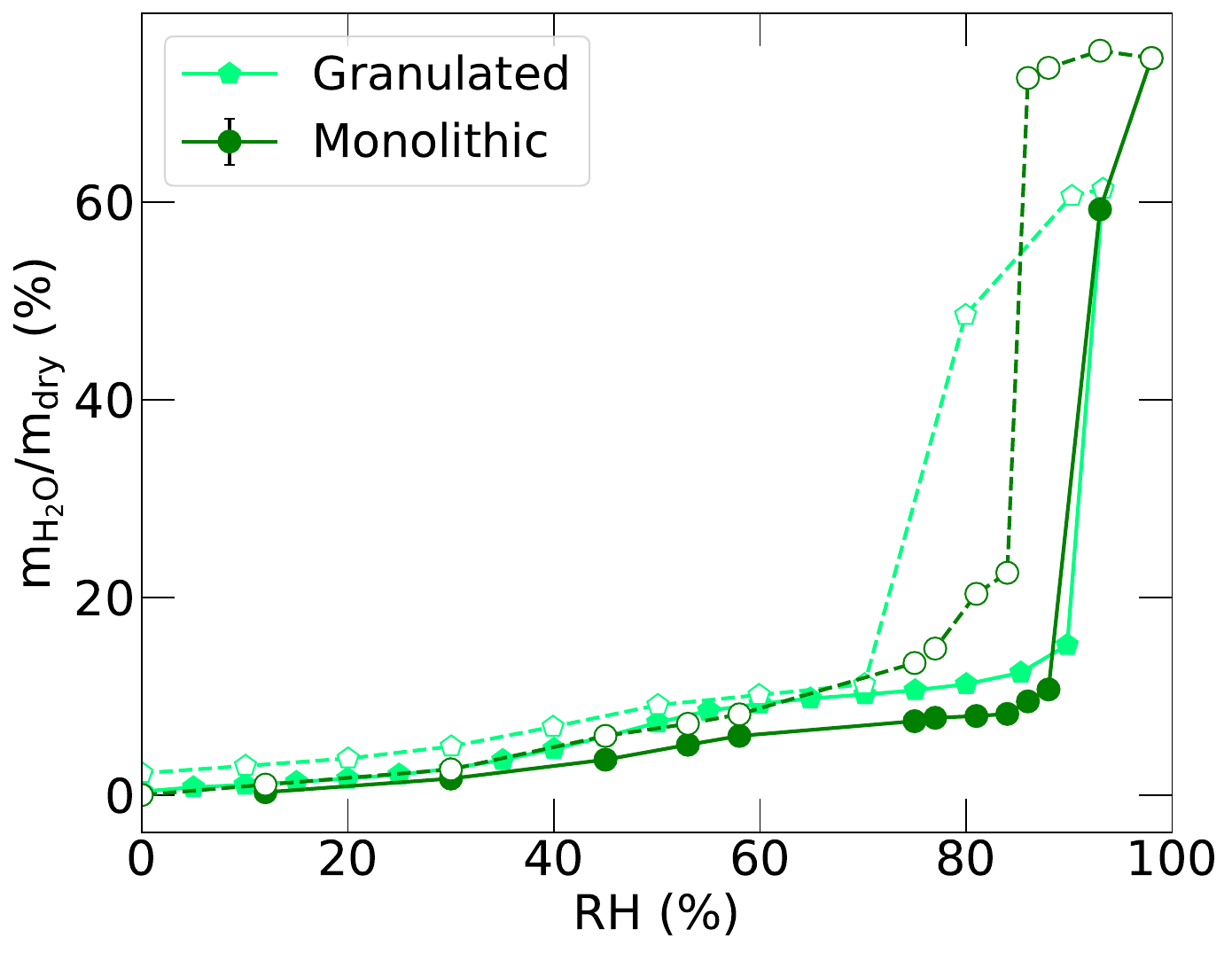} \label{fig:isotherm-mass} }
\subfloat[]{\includegraphics[width=0.5\textwidth]{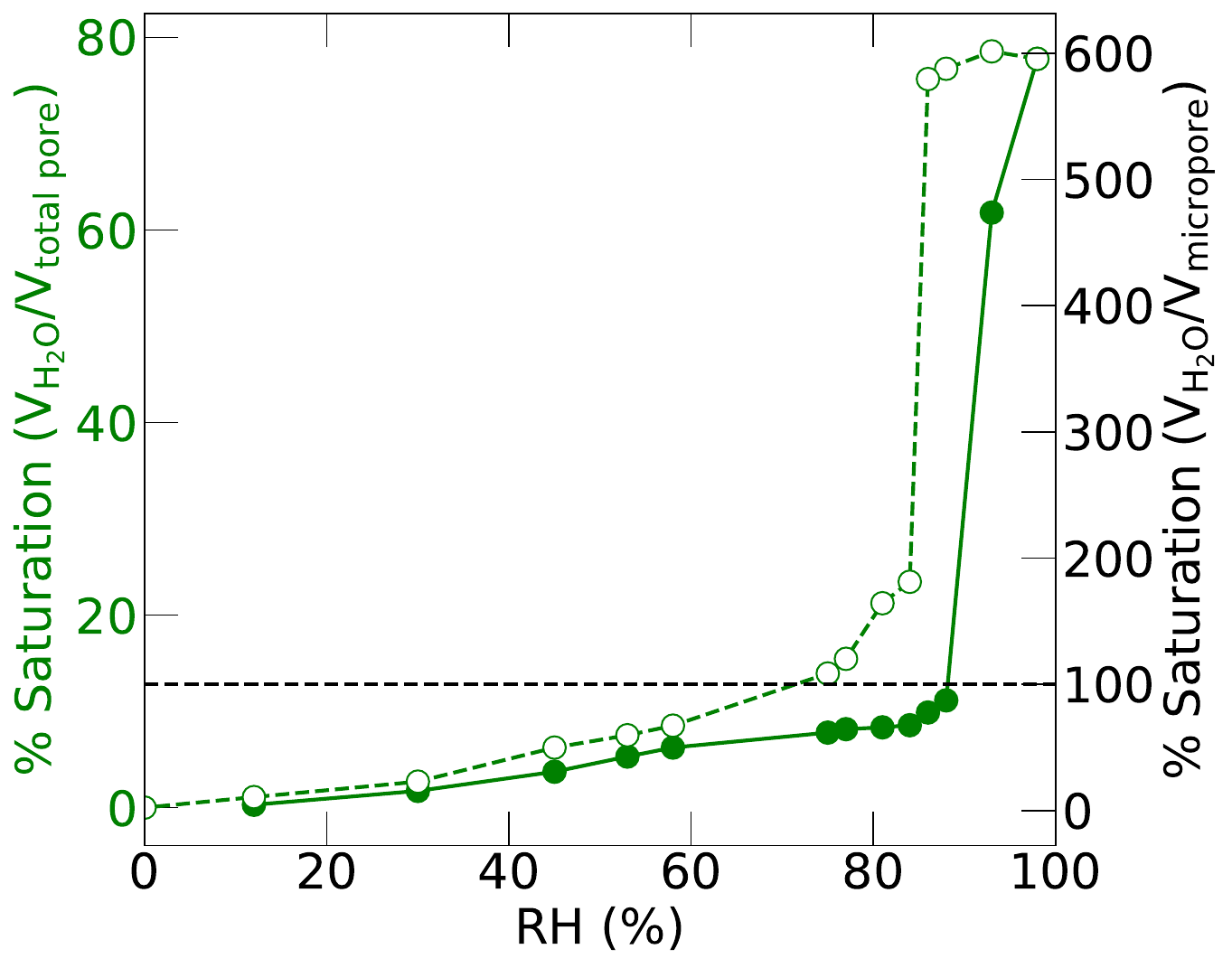}\label{fig:isotherm-saturation}}

    \caption{\textbf{Water sorption isotherms.}(a) Sorption isotherm measured  using the custom made setup for the monolithic sample (green) and sorption isotherm measured using a commercial water sorption balance for the granulated sample (spring green) (error bars represent the 
errors associated with the mass measurements by the balance). (b) Estimated \% sample saturation based on the ratio of the adsorbed water volume to the total pore volume (left scale) and to the micropore volume (right scale) of the monolithic sample. Solid lines with filled markers: adsorption and dashed lines with empty markers: desorption.}
\label{fig:isotherms-water}
\end{figure}

The water sorption isotherm measured for the monolithic xerogel sample by the sorption-ultrasonic setup is shown in Fig.~\ref{fig:isotherm-mass}, along with the isotherm obtained for the granulated xerogel sample with a commercial water sorption balance, as described in the ``Material characterization'' section. Below 70\% RH, both isotherms show type V character~\cite{Thommes2015} that corresponds to water adsorption on a weakly interacting nanoporous material. Thus, water vapor condenses in the micropores first and this process continues with the increase of RH until most of the micropore volume is filled at $\sim$ 70\% RH, and the isotherms start to level off. When RH exceeds $90\%$, mesopores begin to fill, i.e. water is adsorbed in the interparticle spaces. Both monolithic and granulated samples do not reach the full saturation, and the observed discrepancy between the maximum saturation levels for each sample will be discussed in the ``Water adsorption by nanoporous carbon'' section. Desorption occurs following a type H1 hysteresis loop,~\cite{Thommes2015} which arises due to water-carbon interactions coupling with surface heterogeneity and confined geometry effects. At low RH, isotherm loop opening is observed for both granulated and monolithic samples. Fig.~\ref{fig:isotherm-saturation} shows the sample saturation with water, defined as percentage of volume of water adsorbed with respect to both the total pore volume, $V_{\rm pores}$ (left axis) and total micropore volume, $V_{\rm micropore}$ (right axis). Sample saturation with respect to the micropore volume reaches 100\%  at the RH when the mesopore volume filling begins. However, the total sample saturation reaches only $\sim$ 78\%, despite the sample mass change reaching a plateau at 98\% RH (Fig.~\ref{fig:RH93&98-time&mass}), indicating that the mesopores are not filled fully on the timescale of our experiment. 

\subsection{Elastic properties of the xerogel-water composite}
\subsubsection{Waveforms}
\begin{figure}[H]
\centering
\subfloat[]{\includegraphics[width=0.4\textwidth]{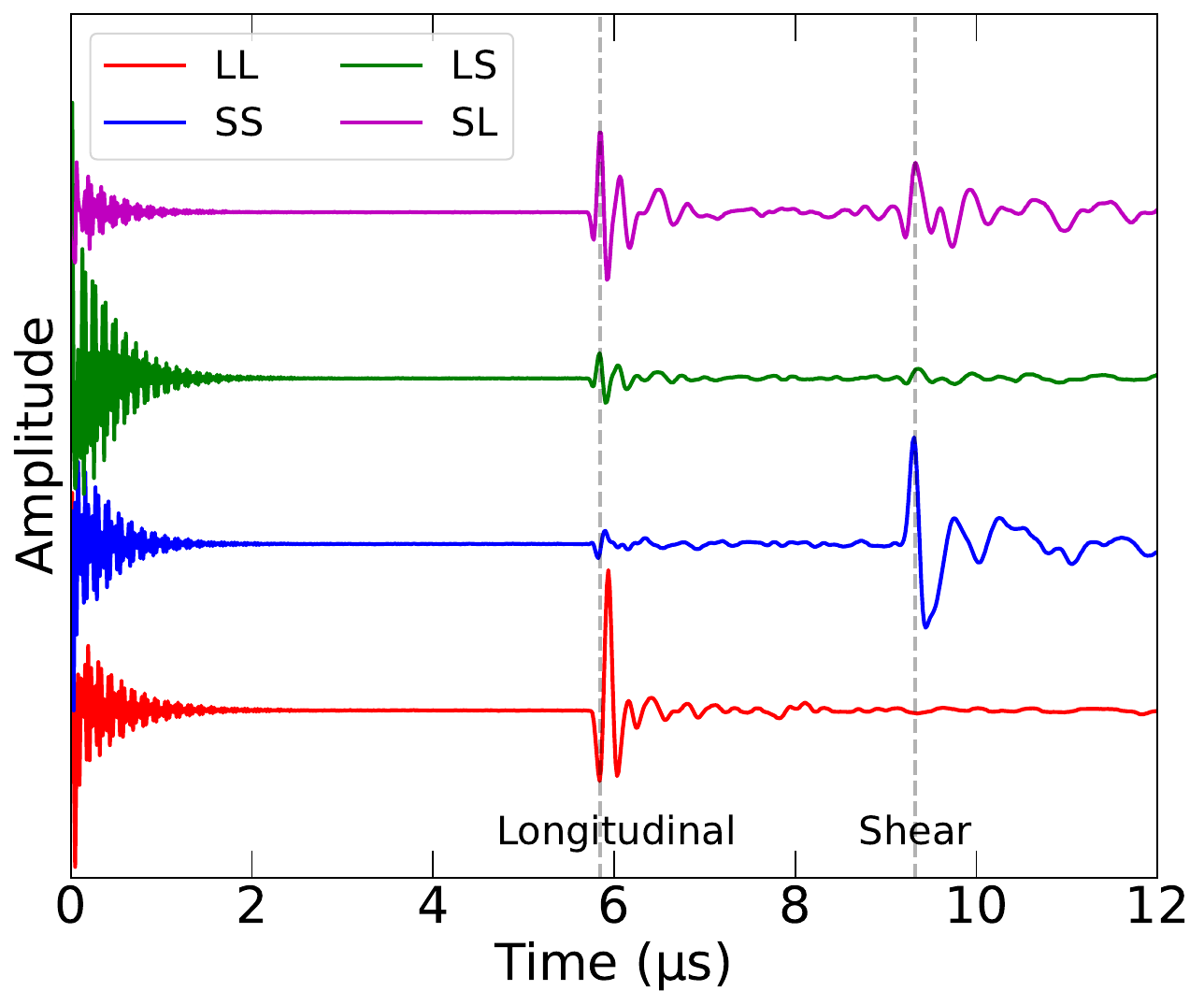} \label{fig:waveform-LS} }
\subfloat[]{\includegraphics[width=0.4\textwidth]{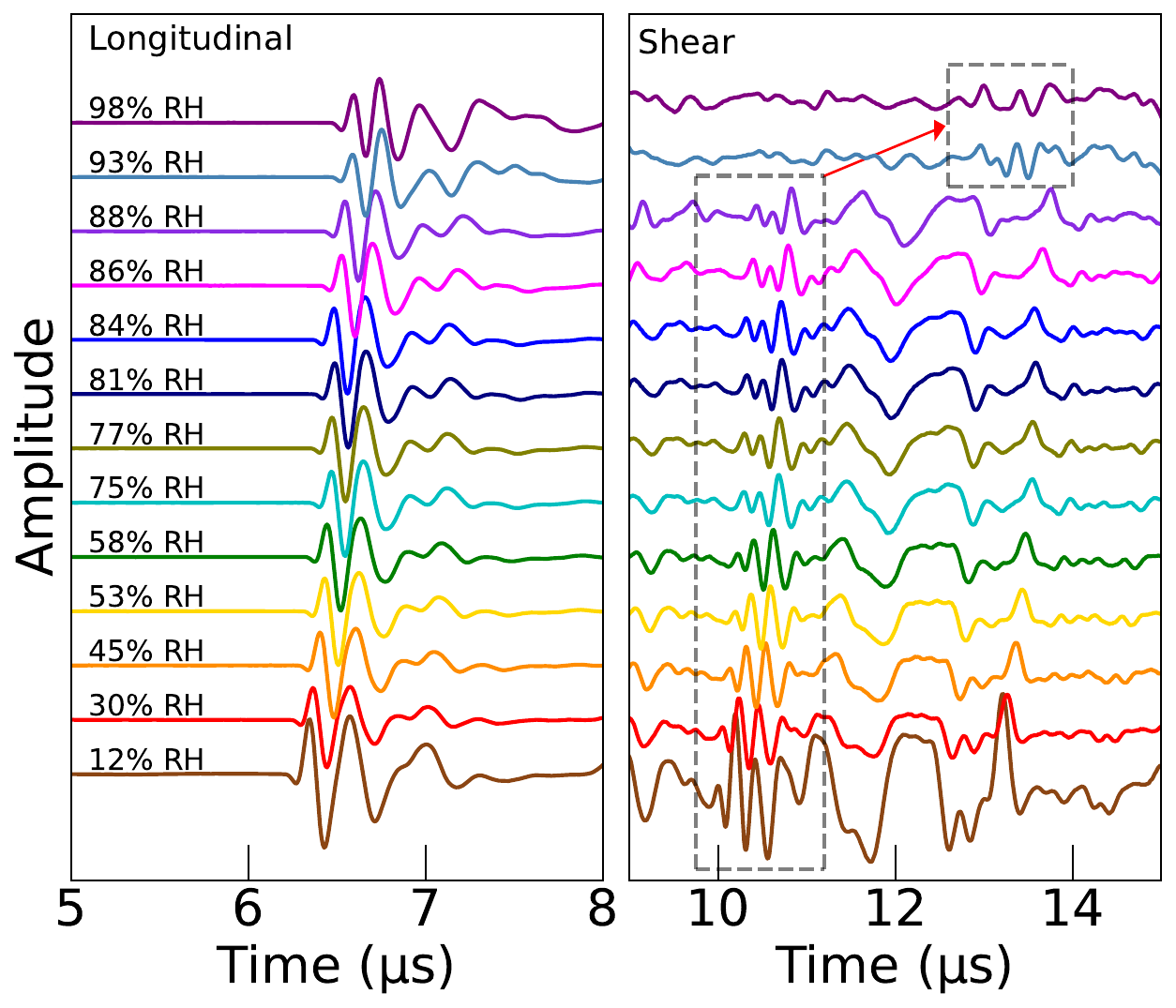}\label{fig:waveform-all RH}}

\subfloat[]{\includegraphics[width=0.40\textwidth]{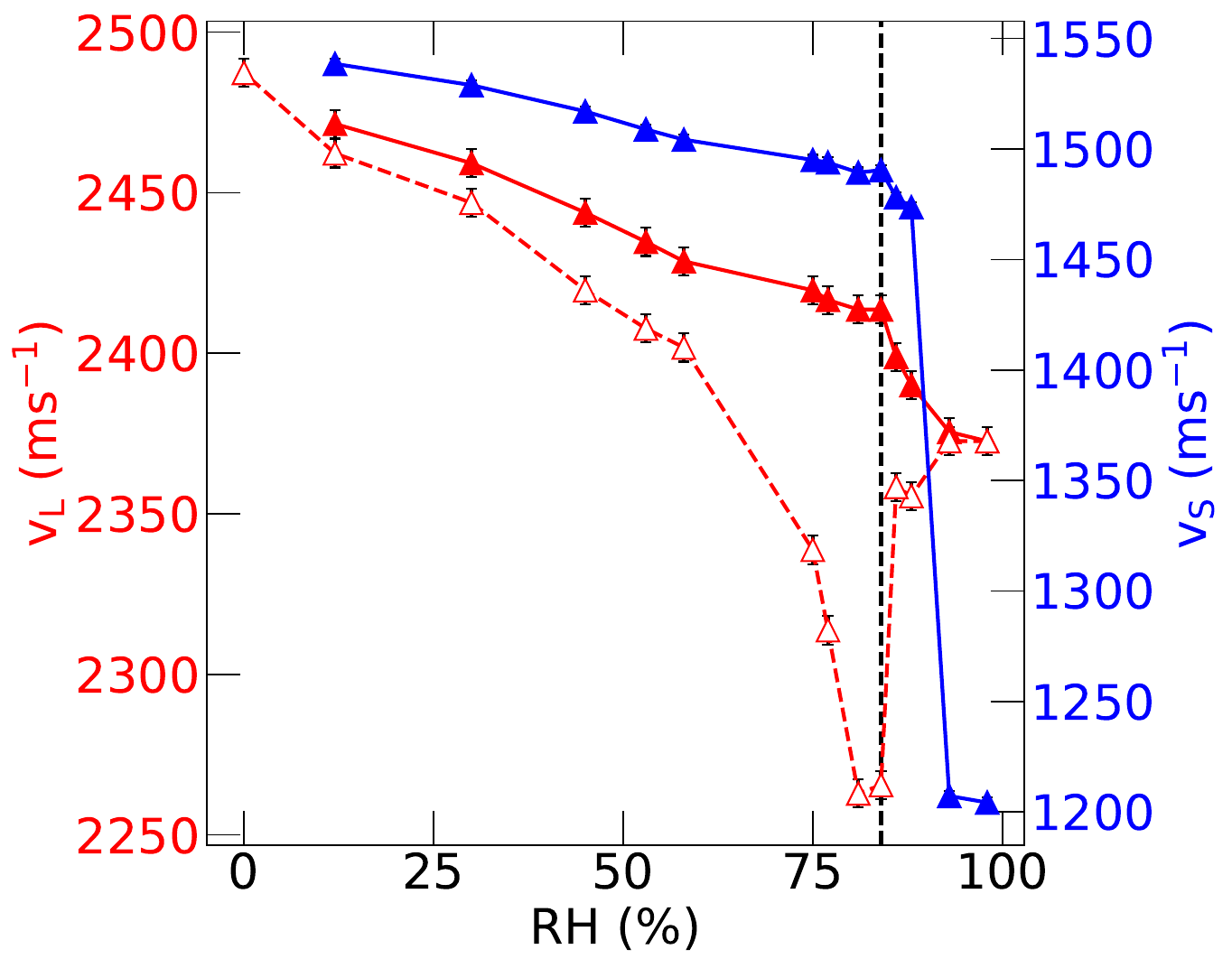}\label{fig:speeds-pressure} }
\subfloat[]{\includegraphics[width=0.4\textwidth]{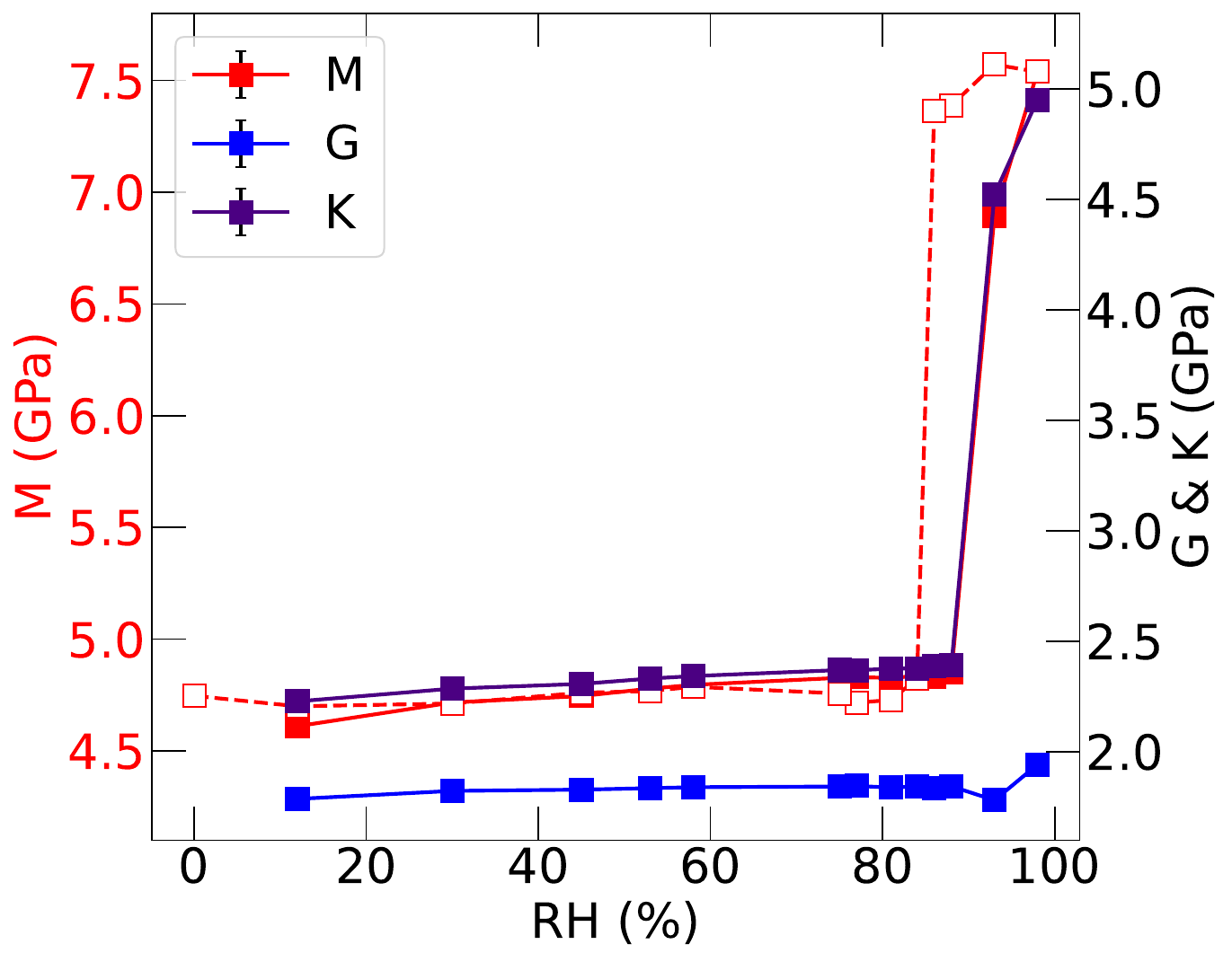} \label{fig:moduli-pressure}}

\subfloat[]{\includegraphics[width=0.4\textwidth]{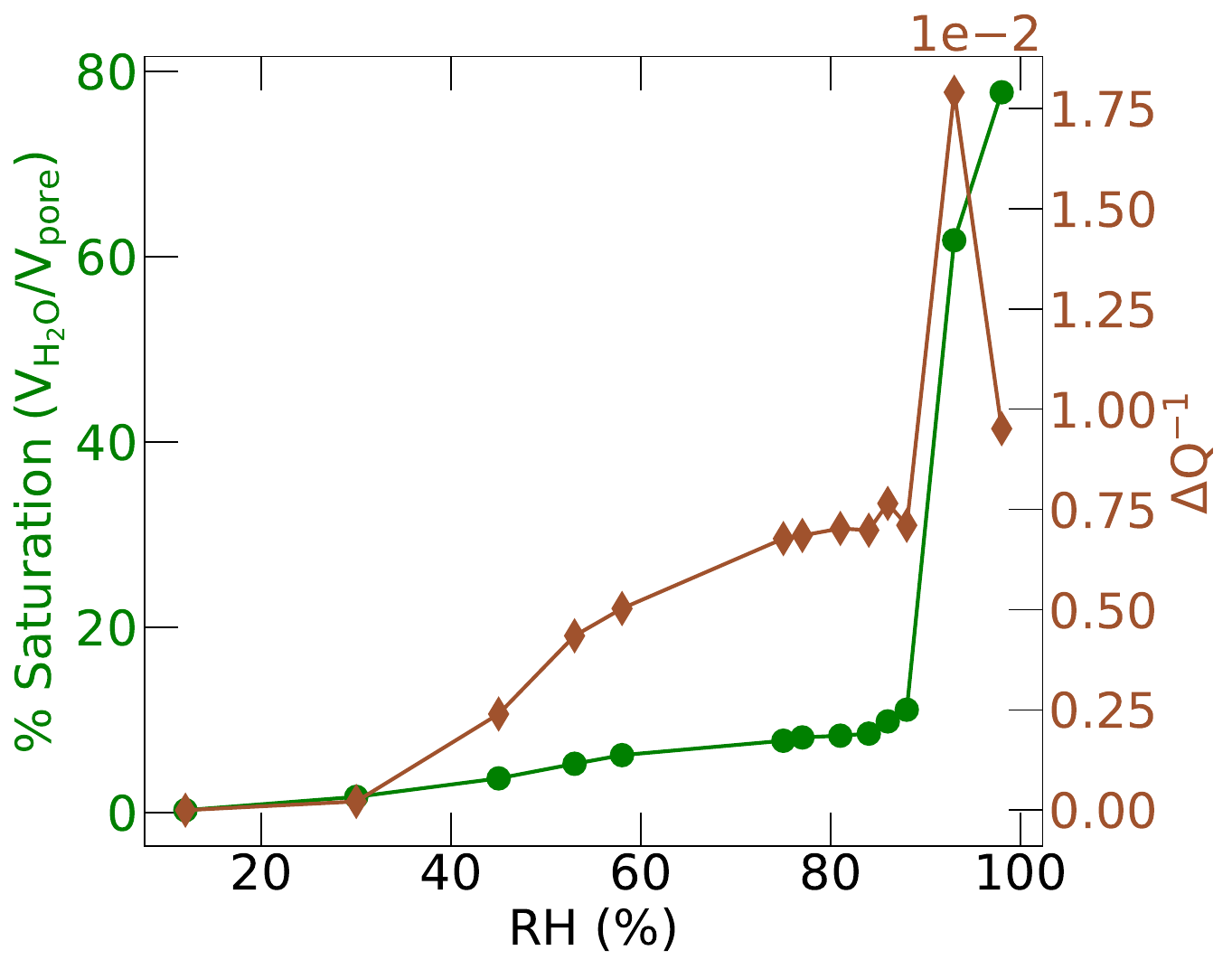}\label{fig:attenuation-pressure} }
\subfloat[]{\includegraphics[width=0.4\textwidth]{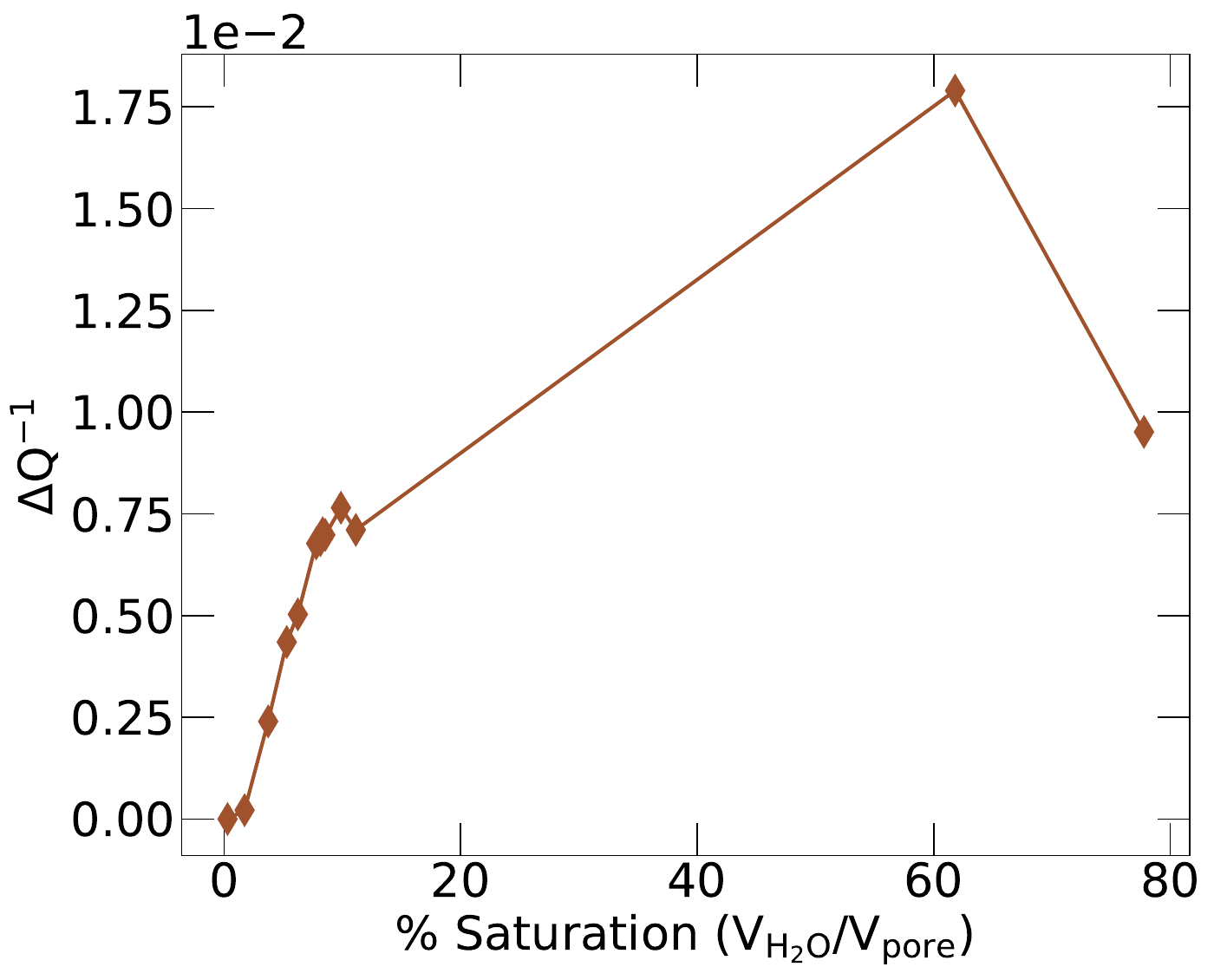} \label{fig:attenuation-saturation}}

\caption{\textbf{Ultrasonic waveforms and sample elastic properties.} (a) Waveforms measured using different XY transducer combinations, where X is the transmitter and Y is the receiver, representing L and S, longitudinal and shear transducers, respectively. (b) Waveforms measured using SL combination at each RH during the adsorption route. (c) Variations in longitudinal and shear wave speeds, where the error bars represent the errors associated with the measurements of the sample length and the travel time. (d) Moduli obtained during water adsorption (solid lines with filled markers) and desorption (dashed lines with empty markers). Attenuation of the longitudinal wave propagation through the sample is shown as a function of (e) RH and (f) sample saturation.}
    \label{fig:waveform}
\end{figure}

In our experiments, it was necessary to record both the longitudinal and shear waves at each point of the adsorption isotherm using the same set of transducers. We could not swap longitudinal and shear transducers during measurements because this would have affected their alignment, introducing time and amplitude artifacts in the measured waveforms. To determine which combination of transducers was the best at discriminating between longitudinal and shear waves in a single measurement, we explored the use of four different transducer combinations, LL, SS, LS, and SL, as shown in Fig.~\ref{fig:waveform-LS}. In these two-letter combinations (L is for longitudinal and S is for shear), the first letter in a sequence corresponds to the transmitter and second to the receiver. We found that the SL transducer combination provided the best discrimination that is consistent with the conclusion by Yurikov et al.~\cite{Yurikov2019}, and so this combination was used to measure the waveforms in water sorption experiments. The waveforms recorded after reaching an equilibrium at each RH level are shown in Fig.~\ref{fig:waveform-all RH}. The travel times for both longitudinal and shear waves were calculated relative to the reference waveform (waveform measured for transducer-couplant-transducer system without sample) (Fig.~\ref{fig:waveform-time}), and those travel times were used to calculate the respective wave speeds.

\subsubsection{Wave speeds, elastic moduli and ultrasonic attenuation}

The variations of both longitudinal, $v_{\rm L}$ and shear, $v_{\rm S}$ wave speeds with RH are shown in Fig.~\ref{fig:speeds-pressure}. During adsorption, both longitudinal and shear waves exhibit a speed decrease, consisting of two distinct regions. In the lower RH region ($12\% \leq \rm RH \leq 84\%$), $v_{\rm L}$ and $v_{\rm S}$ decrease nearly linearly by $\sim 58$ and $\sim 48~\rm m~s^{-1}$, respectively, over a 72\% $\Delta$RH span. Then in the range of $84\%\leq \rm RH\leq 93\%$, both $v_{\rm L}$ and $v_{\rm S}$ follow a much steeper decrease, $\sim 41$ and $\sim 280~\rm m~s^{-1}$ over a 14\% $\Delta$RH span, and reach a plateau at 98\% RH. The initial region in the wave speed plot (Fig.~\ref{fig:speeds-pressure}) corresponds to the initial region in the sorption isotherm (Fig.~\ref{fig:isotherm-mass}), where water is adsorbed mainly in micropores while the mesopores remain empty. Therefore, the initial decrease of the wave speeds at lower RH is related to the micropore filling. Steeper decrease in the speed is seen in the second region above 84\% RH, as the micropore filling is completed and adsorption takes place in mesopores.
When the desorption route is started, the observed sudden drop of longitudinal speed and the complete loss of the shear signal are likely an indication of sample cracking, which was also visually confirmed.

The longitudinal, $M$ and shear, $G$ moduli, calculated from Eqs.~\ref{M-rho} and \ref{G-rho} using the respective wave speeds and the material bulk density adjusted for the adsorbed water mass gain, are shown in Fig.~\ref{fig:moduli-pressure}, using the same vertical span. Both $M$ and $G$ show a gradual slight increase by 5.2\% and 3.2\%, respectively, between 12\% and 88\% RH. Afterwards, at higher RH between 88\% and 98\%, $M$ follows a steep increase by $\sim 55\%$, mimicking the water adsorption while $G$ shows only a slight increase by $\sim 6\%$.

The amplitude of measured waveforms varies significantly with the change in RH (Fig.~\ref{fig:waveform-all RH}) and this is attributed to the ultrasonic attenuation. Ultrasonic attenuation, a measure of the acoustic energy loss per one wave period during the wave propagation through a medium, can be quantified by the inverse of the quality factor ($Q^{-1}$). In this study, spectral ratio method~\cite{Sears1981} which compares the amplitudes of the waveform in frequency domain with the low loss reference waveform was used to determine the ultrasonic attenuation as a function of water saturation. Fig.~\ref{fig:attenuation-pressure} shows the measured change of attenuation ($\Delta Q^{-1}$) of the longitudinal wave during the adsorption process with respect to the attenuation of the waveform measured at 12\% RH (a reference waveform). Since the sample cracked on desorption, and the shape of the waveform changed, the attenuation data is not reported for the desorption process. At a low RH ($<30\%$) the attenuation remains almost constant, but it continuously increases over the range $30\%<\rm RH<88\%$, exhibiting a sharp peak at $88\%<\rm RH<98\%$. Fig.~\ref{fig:attenuation-saturation} shows the attenuation as a function of the sample saturation.

\section{Discussion}

\subsection{Water adsorption by nanoporous carbon}
Based on the production process and chemical and physical post-treatments, carbon materials own their surface chemistry to various functional groups, such as surface oxides, which promote the adsorption of water~\cite{Boehm1994,Boehm2002}. The oxygens in the surface oxides groups are the primary adsorption centers where the water molecules attach via hydrogen bonds. The concentration of the surface functional groups shows a direct effect on the adsorption capacity~\cite{Emmett1948,Dubinin1982,Barton1983}. At low pressures, water adsorption in micropores is initiated by attaching water molecules to the functional groups and as the pressure increases further water adsorption occurs on top of the adsorbed water molecules, growing water clusters near the functional groups. Indeed, for the range of $12\% < \rm RH < 60\%$, the measured isotherm follows type III character describing the water clustering process. As shown in Fig.~\ref{fig:isotherm-mass}, water uptake by the granulated sample is slightly higher (by $\sim 3\%$) than that of monolithic sample at $\rm RH<70\%$. This higher initial uptake on the granulated sample could be because it was initially degassed at $300~\si{\degreeCelsius}$ and only \ch{N2} was used as the carrier gas in the commercial water adsorption balance. The monolithic xerogel sample was not degassed by heating and hence may have contained a small amount of adsorbed water. Additionally, during the experiment it was exposed to water vapor in ambient air, possibly resulting in the presence of co-adsorbed \ch{CO2}, reducing the micropore volume accessible for water filling. Hence the maximum volume adsorbed in the micropore range of the isotherm obtained on the monolithic sample is below the value measured on the granulated sample. However, at highest RH the observed discrepancy of maximum saturation ($\sim 80\%$ vs $\sim$ 60\% ) of monolithic and granulated samples is the other way around. This is mainly due to the much longer exposure time in the case of the monolithic sample. Additionally, the expected water saturation may deviate from the value calculated using the pore volume measured from \ch{N2} isotherms due to the difference in packing of \ch{H2O} and \ch{N2} molecules in the pore space. Adsorption of other molecules from the atmosphere such as \ch{CO2} may also lower the saturation level from 100\%. During water adsorption, the surface chemistry of the pore walls can be irreversibly modified by the water interaction with the surface functional groups (chemisorption), and it may cause the isotherm loop opening observed at low RH.

\subsection{Adsorption-induced deformation}
Sample cracking observed in our experiments during water desorption indicates the occurrence of a plastic deformation in the carbon xerogel, as it takes up and releases water. Notably, although the water sorption isotherm was not affected during cracking, the shape, amplitude, and speed of ultrasonic waveforms changed significantly. Adsorption-induced deformation, i.e. structural change of the solid adsorbent during the fluid adsorption is a widely studied effect in nanoporous materials~\cite{Gor2017review}. During adsorption, fluid exerts pressure on the solid by varying surface stress or change in the solvation pressure inside the pores, and this pressure leads to the deformation of the solid phase. Balzer et al. reported contraction of microporous synthetic carbons during the initial adsorption of \ce{N2}, \ce{Ar}, \ce{CO2} and \ce{H2O} in micropores, followed by expansion as the micropores became fully filled~\cite{Balzer2015xerogels}. The magnitude of contraction and expansion in that study depended on the adsorbate, with water showing a significant contraction. At the initial water adsorption in micropores at low pressures, the formation of a water cluster bridge between the pore walls leads to the material contraction, while the material expansion takes places with the further increase of fluid pressure. This phenomenon of nonmonotonic deformation can be further explained by the positive and negative solvation pressure determined by the pore size and molecular packing. 

Typically, the deformation of porous carbons and glasses due to adsorption is elastic and the measured strain is in the range of $10^{-3} - 10^{-4}$~\cite{Amberg1952, Schappert2014, Balzer2011, Ludescher2021}. However, cracking can occur when local strains exceed the fracture limit, similar to another nanostructured carbon materials--fractal soot particles which restructure upon vapor condensation. Soot particles are aggregates of primary carbon spherules chemically fused together by carbon necks, and their lacey fractal morphology becomes compact due to the capillary forces exerted by liquid condensates~\cite{Chen2018} that induce the fracturing of carbon necks. In a recent study~\cite{Karunarathne2025}, using our discrete element method (DEM) model~\cite{Demidov2024} we have shown that that shear stress is much higher (by three orders) than tensile stress due to a combination of torques and levers in a fractal aggregate, and most of that stress is focused on a few necks, resulting in their fracture. A similar process may have occurred in the xerogel sample, producing a single long crack.

\subsection{Estimation of water bulk modulus}

The elastic modulus of the composite xerogel-water sample can be used to retrieve the modulus of confined water, given known values of moduli of the solid carbon and the porous sample. The poroelastic theories proposed by Gassmann~\cite{Gassmann1951,Berryman1999}  and Biot~\cite{Biot1956i,Biot1956ii} describe the elastic properties of fluid-saturated porous medium and their dependence on the elasticity of the pore fluid. The Gassmann equation relates the  effective bulk modulus of the fluid--saturated porous medium to its porosity $\phi$ and the bulk moduli of the dry porous medium $K_{0}$, solid phase $K_{\rm s}$ and the fluid $K_{\rm f}$ as,

\begin{equation}
\label{Gassmann}
    K_{\rm G}=K_{0}+\frac{\alpha^{2}}{\frac{(\alpha-\phi)}{K_{\rm s}}+\frac{\phi}{K_{\rm f}}}
\end{equation}
where the Biot-Willis coefficient $\alpha = 1-\frac{K_{0}}{K_{\rm s}}$. This relation clearly reflects the effect of the elastic properties of fluid on the effective elastic properties of the fluid-filled porous medium. Previous studies utilized the Gassmann equation to explore the elastic properties of fluids confined in nanoporous media and observed deviation of their elastic properties from those
in the bulk~\cite{Page1995, Gor2018Gassmann, Maximov2018, Sun2019, Ogbebor2023}. The deviation is due to the interplay between the solid-fluid interactions and the pore structure, including the size, shape and connectivity of the pores in the material. These experimental observations are backed by recent density functional theory calculations and molecular simulations~\cite{Gor2014,Dobrzanski2021}. However, since the Gassmann equation was derived for fully saturated porous media, the fluid bulk modulus $K_{\rm f}$ parameter in the equation must be redefined for the partially saturated porous media. 

For a porous medium saturated with a mixture of two fluids distributed uniformly within the pore space, where one of the fluids could be air or vapor, the mixture can be considered as a single fluid with an effective bulk modulus $K_{\rm f}$ given by,
\begin{equation}
\label{GW}
  \frac{1}{K_{\rm f}}=\frac{S_{1}}{K_{\rm f1}}+\frac{S_{2}}{K_{\rm f2}},
\end{equation}
where $S_{1}$ and $S_{2}$ are the volume fractions of each fluid with bulk moduli $K_{\rm f1}$ and $K_{\rm f2}$, respectively. The combination of Eqs.~\ref{Gassmann} and ~\ref{GW} is defined as the Gassmann-Wood (GW) limit~\cite{Johnson2001}. 

\begin{figure}[H]
\centering
\subfloat[]{
\includegraphics[width=0.5\textwidth]{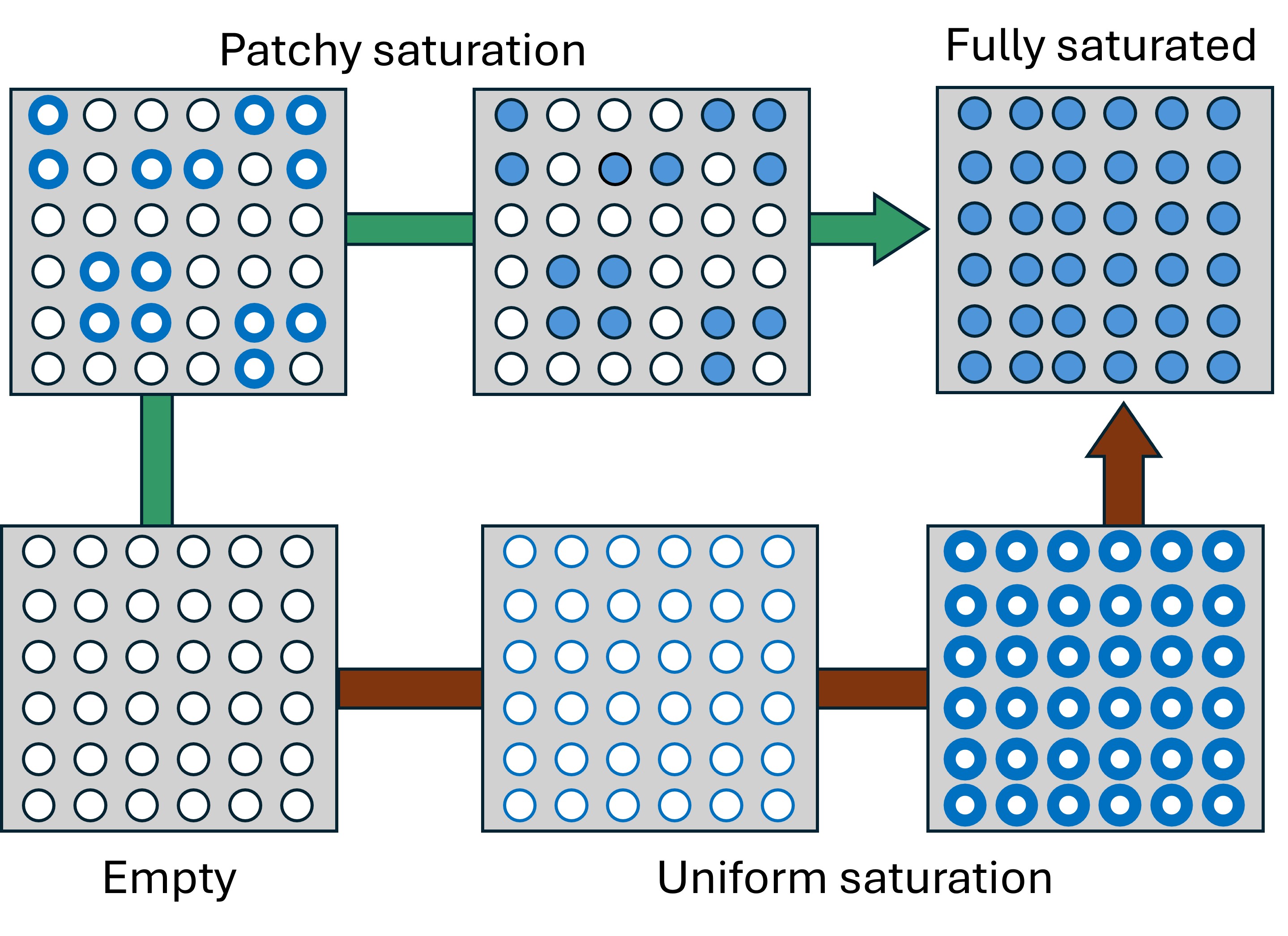} \label{saturation-schematic} }
\subfloat[]{
\includegraphics[width=0.5\textwidth]{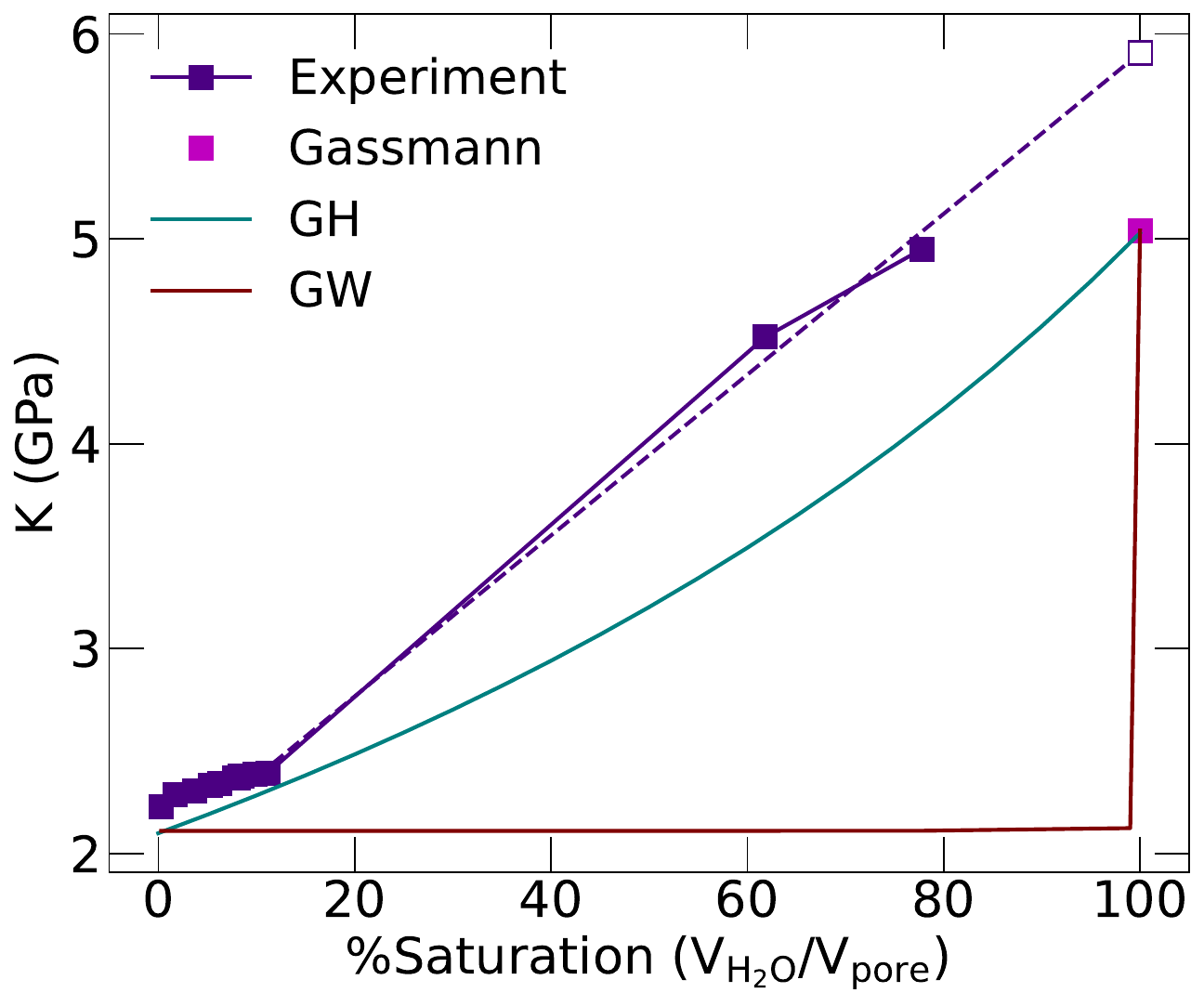} \label{Kgassmann}}

    \caption{\textbf{Estimation of water bulk modulus.} (a) Schematic illustration of the spatial distribution of the fluid-filled pores during the uniform saturation and the patchy saturation. (b) The effective bulk modulus of the xerogel sample, obtained from ultrasonic measurements and estimated from Gassmann, GW and GH approximations (indigo dashed line with an empty marker: extrapolation of experimental data to find the bulk modulus at full saturation).}

\end{figure}

However, if the characteristic size of the pore clusters saturated with two fluids is much larger than the hydraulic diffusion length, the equilibration of fluid pressure between clusters of pores with different levels of saturation is not reached within one time period of the wave. This saturation is known as the patchy saturation where clusters of pores are fully saturated by fluid 1 while the rest of pores are saturated by fluid 2 (Fig.~\ref{saturation-schematic}). The bulk moduli, $K_{\rm G}^{1}$ and $K_{\rm G}^{2}$ of the pore clusters saturated with each fluid can be obtained from the Gassmann equation using the respective fluid bulk moduli, $K_{\rm f1}$ and $K_{\rm f2}$. The effective bulk modulus, $K_{\rm GH}$ of the medium represented by a mixture of pore clusters can be expressed by the volume fraction-weighed longitudinal moduli of each pore cluster, as given by Hill theorem~\cite{Hill1963}.  
\begin{equation}
 \frac{1}{K_{\rm GH}+\frac{4}{3}G_{0}}=\frac{S{1}}{K_{\rm G}^{1}+\frac{4}{3}G_{0}}+\frac{S{2}}{K_{\rm G}^{2}+\frac{4}{3}G_{0}}
\end{equation}
This relation is known as Gassmann-Hill (GH) limit and it considers the shear modulus to be independent of the saturating fluid and equal to the shear modulus of dry porous medium, $G_{0}$. Both Gassmann-Wood and Gassmann-Hill approximations allow to express the dependence of material's effective bulk modulus on the fluid saturation of the sample~\cite{Gurevich2024}. Fig.~\ref{Kgassmann} shows the effective bulk modulus of carbon xerogel plotted against its water saturation as obtained from ultrasonic measurements (indigo) and as estimated using GH (teal) and GW (maroon) approximations. A point obtained by Gassmann equation assuming a 100\% saturation (magenta) is also shown. To obtain these estimations, we used the bulk modulus of dry sample measured in vacuum $K_{\rm 0}=2.11 ~\rm GPa$, bulk modulus of solid phase $K_{\rm s}=48.2 ~\rm GPa$ estimated from the Kuster and Toks\"oz (KT) effective medium theory~\cite{Kuster1974}, water bulk modulus at $20~\si{\degreeCelsius}$ $K_{\rm W}=2.2~ \rm GPa$, and the bulk modulus of water vapor $K_{\rm v}=2.2 ~\rm kPa$~\cite{Wagner2002}. Experimentally measured effective bulk modulus shows a good agreement with GH approximation, where bulk modulus increases with the water saturation. This agreement confirms the spatial heterogeneity of the water filled pore space in xerogel that leads to the patchy saturation. To obtain an estimate of the bulk modulus of the fully saturated xerogel sample, we extrapolated the experimental data as shown in Fig.~\ref{Kgassmann} (indigo dashed line with empty markers). Using this value ($K=5.91 ~\rm GPa$), we estimated the bulk modulus of adsorbed water $K_{\rm W}$ at 100\% saturation by the Gassmann relation (Eq.~\ref{Gassmann}). This estimated $K_{\rm W}$ reaches 2.89 GPa at the full saturation, which is 30\% higher than that of bulk water (2.2 GPa). While the linear extrapolation used to obtain this value should be used for quantitative predictions with caution, the deviation of the bulk modulus of adsorbed water from the bulk water can be due to the interactions between the water molecules and the carbon surface. This result is consistent with the reported positive deviation of bulk modulus of water confined in nanoporous Vycor glass~\cite{Ogbebor2023}. 

Kuster and Toks\"oz (KT) effective medium theory is typically used to estimate the effective bulk, $K$ and shear, $G$ moduli of a porous medium, using the porosity, bulk, $K_{\rm s}$ and shear, $G_{\rm s}$ moduli of the solid phase, and the pore geometry. Here we follow the opposite relying on Kuster and Toks\"oz (KT) effective medium theory derived for cylindrical pore geometry to calculate $K_{\rm s}$ using experimentally measured elastic properties of dry porous material, $M_{0}$ and $G_{0}$, and the porosity, $\phi$ (for more details, see Refs.~\cite{Sun2019, Ogbebor2023}). Since the estimated $K_{\rm s}$ can deviate from the actual value, an estimation of the effect of $K_{\rm s}$ on the calculated materials effective bulk modulus and the water bulk modulus is important. Previous studies have reported the dependence of the bulk modulus of amorphous carbons on the hybridization and the materials density, and the reported bulk moduli range from 80 to 516 GPa over the density variation form 2.4 to 3.3 g/cm$^{3}$~\cite{Jensen2015}. We found that a relative uncertainty in $K_{\rm s}$ by $+50\%$ $(-50\%)$ results in a $+2\%$ $(-6\%)$  uncertainty in $K_{\rm GH}$ and a $-3\%$ $(+12\%)$ uncertainty in the calculated water bulk modulus $K_{\rm W}$.
 
\subsection{Elastic properties of xerogel-water composite}
The presence of bimodal pore size distribution in nanoporous carbons, consisting of micropores and mesopores, typically produces a characteristic two-step water sorption isotherm with initial micropore filling at lower pressures and  mesopore filling at high pressures~\cite{Thommes2013}. In this study, a clear transition from micropore filling to mesopore filling is reflected not only in the water sorption isotherm, but also in the variations of wave speeds, ultrasonic attenuation, and elastic moduli. Fig.~\ref{fig:percent change-mass,speed,moduli} shows the \%change of mass density and elastic properties relative to their values at 12\% RH expressed as a function of RH and \%saturation. In the RH range from 12\% to 88\%, density increases by 12\% while $v_{\rm L}$ and $v_{\rm S}$ decrease by 3.3\% and 4.3\%, respectively. From 12\% RH to 30\% RH (between the first two data points) both longitudinal and shear moduli increase relatively sharply by $\sim 2.5\%$ . This material stiffening can be related to the pre-stress arising during the material contraction at initial micropore filling~\cite{Balzer2015xerogels}. The material contraction can be viewed as a porosity reduction, which leads to the increase of material stiffness. After that, from 30\% RH to 88\% RH the shear modulus remains constant while the bulk modulus increases steadily by another $\sim 4.7\%$ and this is an indication of patchy saturation of water with complete micropore filling. At higher RH (from 88\% to 98\%), the sample gains a significant amount of water with the mesopore filling and the density increases steeply by 65\%. With this steep density increase, $v_{\rm S}$ decreases by 17\% and therefore the shear modulus, $G$ remains constant while the longitudinal and bulk moduli follow a sharp increase further confirming the patchy saturation.  

The increase of the attenuation at RH ($<80\%$) (Fig.~\ref{fig:attenuation-pressure}) points to a spatial heterogeneity of the water-filled pore space, i.e. the complete micropore filling while the mesopores remain empty. The attenuation peak at 93\% RH reflects the highest heterogeneity of the water-filled pore space where almost all the micropore space is completely filled. The decrease of attenuation from 93\% RH to 98\% RH reflects the transition of heterogeneous filling to homogeneous filling as the sample reaches the full saturation with fully filled micropores and mostly filled mesopores. Note that in order to make the discussion more quantitative more RH points are needed, so that the peak is represented by more than a single point. Previous ultrasonic studies of hexane~\cite{Page1995} and water ~\cite{Ogbebor2023} adsorption on Vycor glass related the sharp attenuation peak to the capillary condensation. In mesoporous Vycor glass, both hexane and water adsorption occurs uniformly at lower vapor saturations until the capillary condensation starts to fill the pores completely. At the onset of the capillary condensation, adsorption becomes non-uniform as the sample contains both completely and partially filled pores, and later it becomes uniform again with the full saturation of all the pore volume. The attenuation is attributed to the relaxation of fluid pressure causing viscous dissipation as the wave propagates through the sample with different levels of liquid saturation. Hence, our study demonstrates the utilization of acoustic attenuation measurements to evaluate the spatial distribution of the water-filled pore space within the overall sample volume.

\begin{figure}[H]
\centering
\subfloat[]{\includegraphics[width=0.5\textwidth]{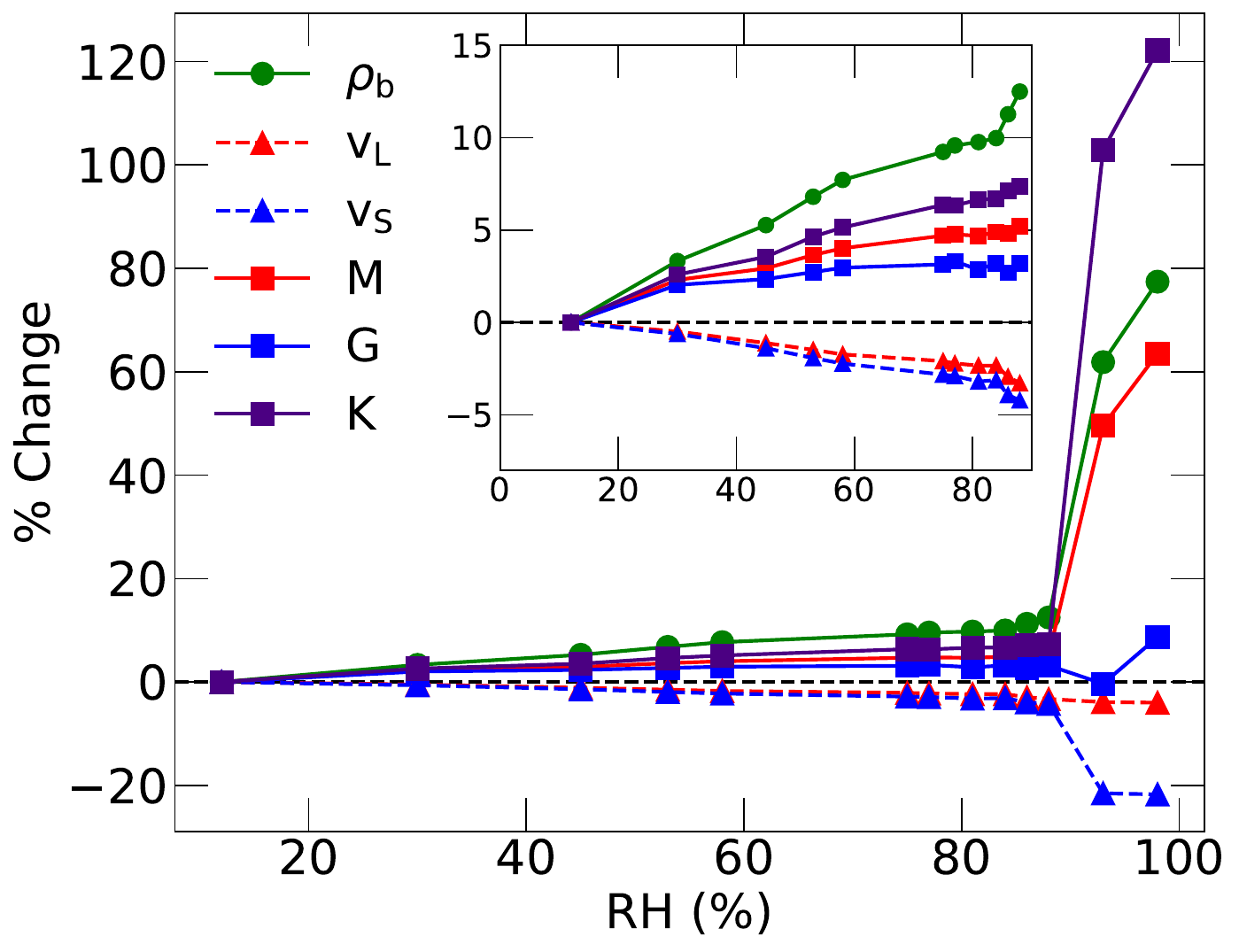}\label{change-pressure} }
\subfloat[]{\includegraphics[width=0.50\textwidth]{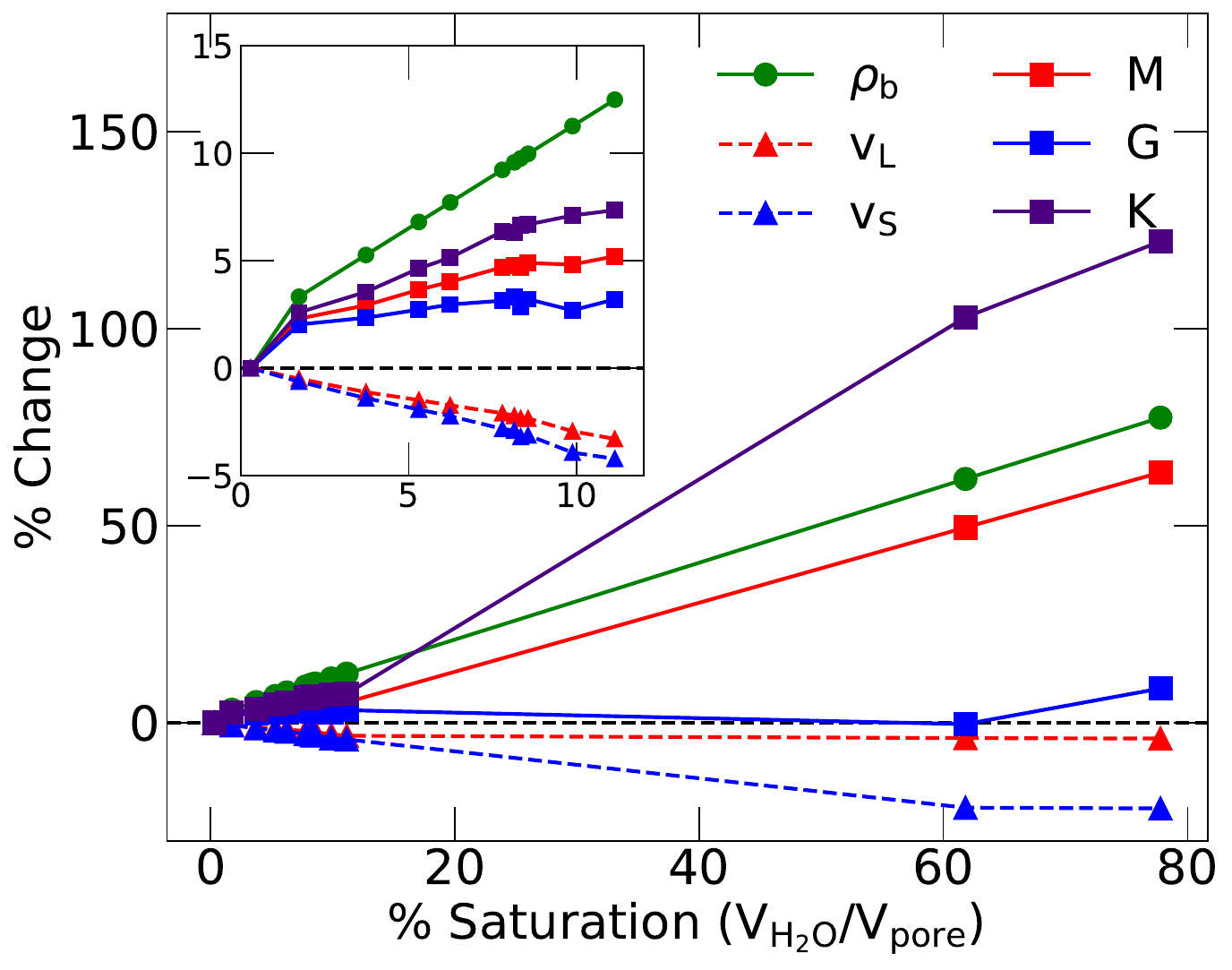} \label{change-saturation}}

    \caption{\textbf{Relative variations of mass density and elastic properties}. Percent changes of mass density, wave speeds and elastic moduli relative to their values at 12\% RH as a function of (a) RH and (b) sample saturation. The insets show the zoomed regions below 88\% RH and 11\% saturation, respectively} 
    \label{fig:percent change-mass,speed,moduli}
\end{figure}

\section{Conclusion}

We relied on the ultrasonic wave propagation characteristics, such as wave speed and amplitude to retrieve the sample elastic modulus and wave attenuation, gaining a deeper understanding of the water adsorption and filling mechanism of nanoporous carbon xerogel. To achieve this, the ultrasonic wave speeds were measured concurrently with water sorption isotherm using a novel adsorption-ultrasonic experimental setup. Overall, water sorption follows type V isotherm with type H1 hysteresis loop. The observed distinct trends in the variation of wave speeds and elastic moduli demonstrate a clear correlation with the density variation obtained from the adsorption isotherm. In the low RH region, sorption follows type III isotherm corresponding to the clustering of water molecules and micropore filling. When the micropores are filled, water continues to adsorb in mesopores, suggesting a two step filling mechanism. The observed variation in acoustic attenuation with the increase in water saturation reflects the transformation from  spatially heterogeneous towards spatially homogeneous fluid-filled pore space, as the adsorption proceeds from micropore to mesopore filling. Using Gassmann-Hill relation, we estimated the bulk moduli of water confined in micropores and mesopores that reflect the configuration of water molecules inside those pores and the interaction of water molecules with carbon surface. Overall, water adsorption is governed by both the surface chemistry and the pore size distribution of the carbon xerogel. Our study thus demonstrates utilization of ultrasonic testing as a nondestructive tool for probing the fluid adsorption mechanism in nanoporous media, and overall characterization of porous materials using acoustical techniques~\cite{Horoshenkov2017}.

\section{Materials and Methods}
\subsection{Carbon xerogel synthesis}

\begin{figure}[H]
\centering
\subfloat[]{\includegraphics[width=0.65\textwidth]{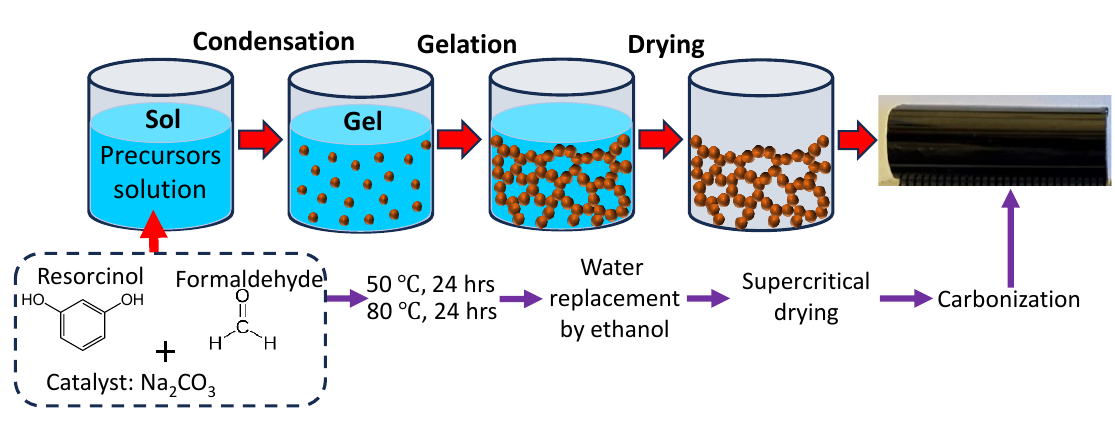} \label{fig:synthesis-xerogel}}
~
\subfloat[]{\includegraphics[width=0.35\textwidth]{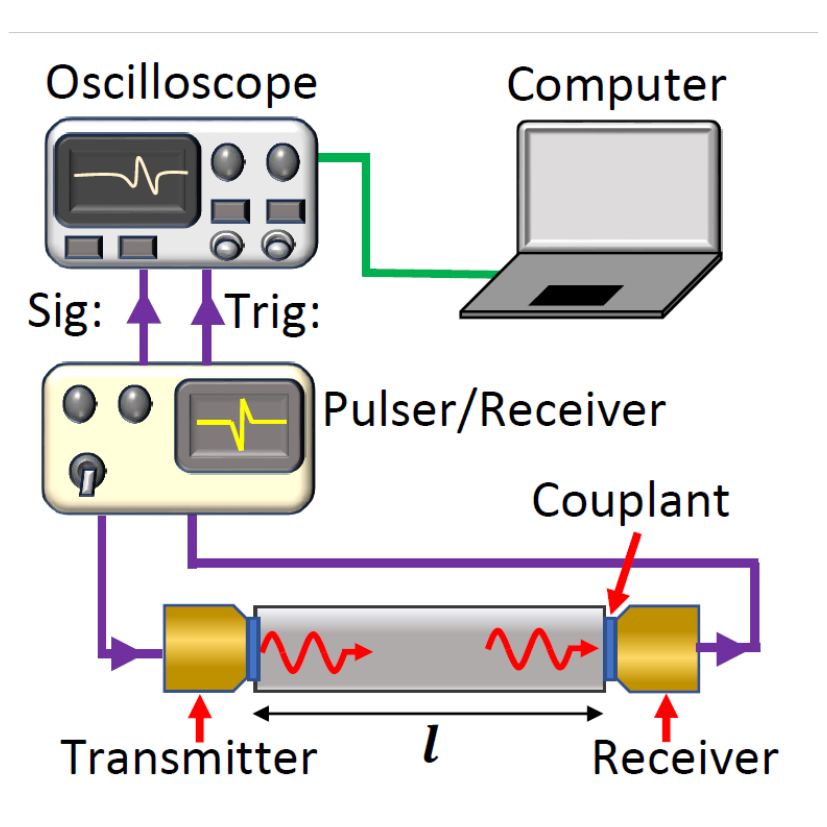}\label{fig:setup-pulse transmission}}

\subfloat[]{\includegraphics[width=0.35\textwidth]{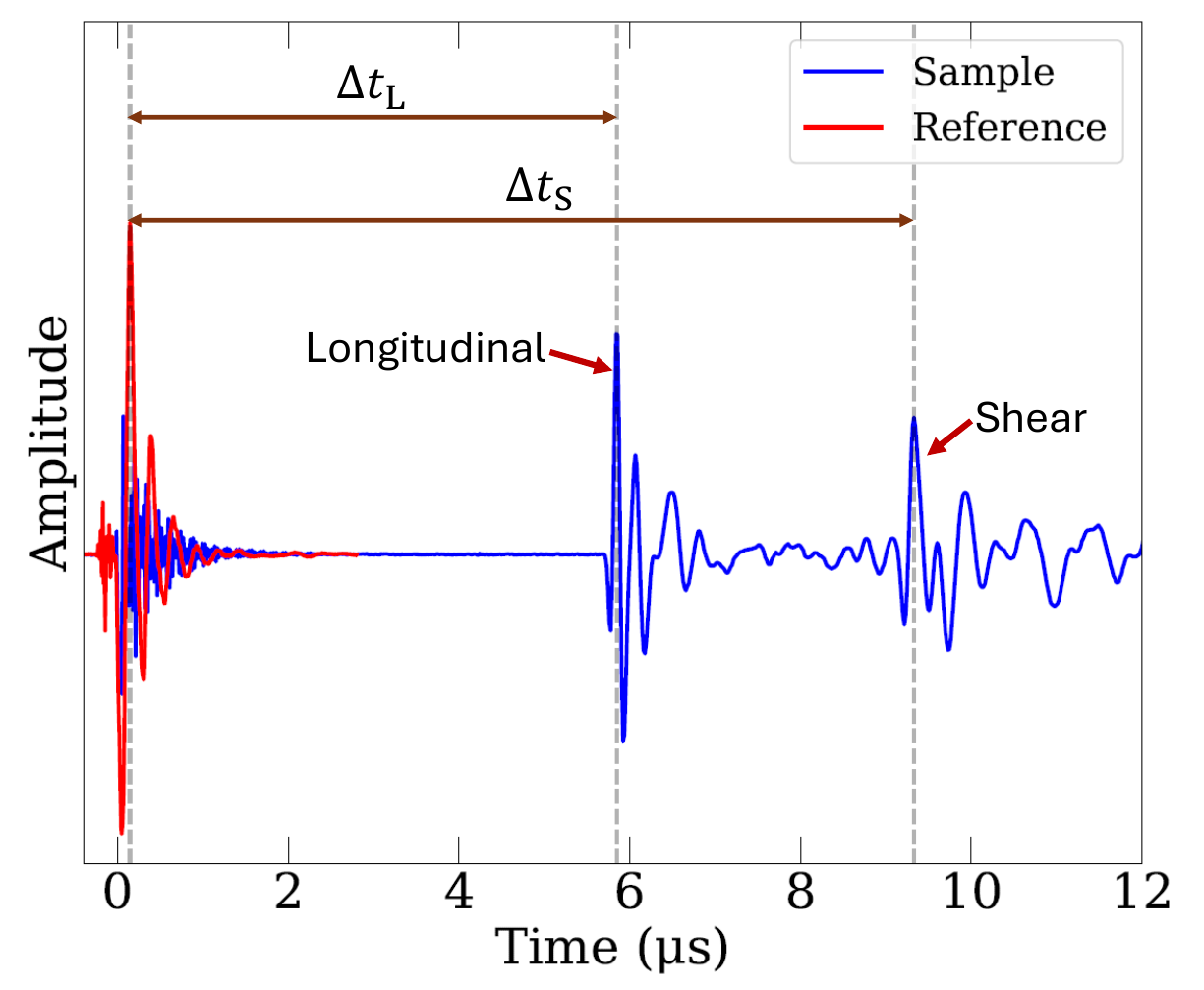}\label{fig:waveform-time}}~
\subfloat[]{\includegraphics[width=0.65\textwidth]{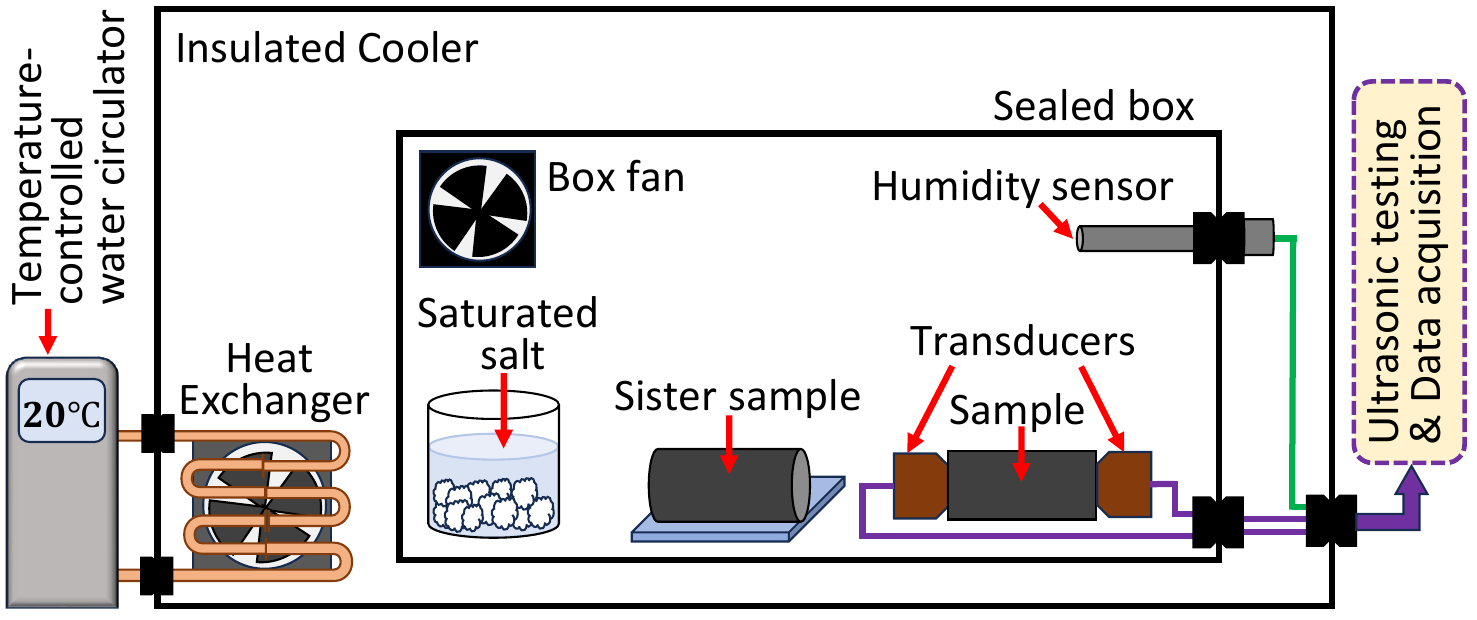} 
    \label{fig:setup-adsorption-ultrasonic}}

\caption{\textbf{Experimental methods.}(a) Schematic illustration of carbon xerogel synthesis procedure. Precursors solution (sol) is allowed to undergo polycondensation by heat treatment to form gel. The water medium is then replaced by ethanol and followed by supercritical drying with carbon dioxide to remove the liquid phase. The dried sample is then carbonized by pyrolysis. (b) Schematic of the experimental setup for the ultrasonic pulse transmission method. (c) A typical waveform measured for the carbon xerogel sample and the determination of the time of flight relative to the reference waveform. (d) Schematic of adsorption-ultrasonic experimental setup.}
    \label{fig:materials and methods}
\end{figure}

Carbon xerogel samples were synthesized by pyrolysis of
organic precursors (Fig.~\ref{fig:synthesis-xerogel})~\cite{Pekala1989,Wiener2004}. The carbon xerogel precursor was prepared by polycondensation reaction from a mixture of resorcinol (1,3-dihydroxybenzene) and formaldehyde (1:2 molar ratio) diluted by deionized water. Sodium carbonate
was added as a catalyst. The solution was poured into analytical glass vials (1.6 cm in diameter) and sealed airtight using screw caps. Then, the solution was exposed to $50~\si{\degreeCelsius}$ for 24 hrs and $85~\si{\degreeCelsius}$ for 24 hrs for gelation and curing. After the gelation  step, the water in pores of wet gel was replaced by ethanol by a solvent exchange step and the samples were dried under supercritical conditions (\ce{CO2}, $45~\si{\degreeCelsius}$, 100 bar, 24 hrs). The dry samples were then carbonized at $900~\si{\degreeCelsius}$ for 60 minutes under argon atmosphere (heating rate was 3 K/min).

\subsection{Material characterization}
The microstructure of the synthesized samples was imaged by scanning electron microscope (SEM). The  bulk (apparent) density, $\rho_{\rm b}$ was obtained from the volume and mass measurements of a cylindrical sample. The \ce{N2} and \ce{CO2} sorption data were obtained using a volumetric sorption analyzer (ASAP 2020 from Micromeritics Inc., USA). Before analysis the sample was degassed under vacuum at $300~\si{\degreeCelsius}$ for at least 5 hours. Evaluation of the BET surface from the \ce{N2} sorption data was performed following the IUPAC technical report to select the correct pressure range for this microporous material.~\cite{Thommes2015} Determination of pore size distribution (PSD) was carried out using Micromeritics Microactive software. For the PSD of the mesopores, a non local density functional theory (NLDFT) Kernel for nitrogen on carbon with cylindrical pore geometry was used (\ce{N2}-2D-NLDFT). The micropore distribution was obtained using a combination of two NLDFT models for \ce{CO2} and \ce{N2} sorption measurements (\ce{CO2} at 273~K in carbon and \ce{N2} at 77~K on carbon slit pores) using the “NLDFT advanced PSD” tool of Microactive.\cite{Jagiello2015}

Before conducting ultrasound-adsorption experiments, water sorption isotherm was obtained using a commercial sorption balance (SPS11-10$\rm \mu$ from ProUmid GmbH  Co. KG, Germany). The device uses a humidified carrier gas stream (nitrogen) to expose the sample to a defined relative humidity at constant temperature. The humidity is measured by a calibrated capacitive humidity sensor. The amount of water adsorbed is determined from the sample mass gain. Before the measurement, the sample was crushed into granules of about 3 mm size to accelerate the sorption process. Further we refer to this sample as ``granulated sample". The measurement was performed at $25~\si{\degreeCelsius}$ in the relative humidity range from 0\% to 95\% RH with steps of 5\% RH for adsorption an desorption. As a condition for equilibrium of each step, the mass change of the sample had to be lower than 0.01\% per 130 minutes. To determine the mass of the empty sample the granules were pre-dried in an oven under vacuum and subsequently conditioned at $50~\si{\degreeCelsius}$ and 0\% RH inside the sorption balance chamber.

\subsection{Ultrasonic measurements}
Ultrasonic pulse transmission technique was used to obtain both longitudinal, $v_{\rm L}$ and shear, $v_{\rm S}$ sound speeds through the monolithic xerogel sample~\cite{Asay1969,Brown1995,Kohlhauser2013}. In this method, propagation of ultrasonic waves is observed by transmitting an ultrasonic pulse from a transducer (transmitter) attached to one end of the test sample and receiving the propagated pulse by the transducer (receiver) attached to the opposite end of the sample (Fig.~\ref{fig:setup-pulse transmission}). For porous materials, the use of conventional fluid couplants is limited as they tend to diffuse into pores of the material. Therefore, thin nitrile rubber sheet was used as the coupling medium between the sample and transducers, following Ogbebor et al.~\cite{Ogbebor2023}. The time for the ultrasonic wave propagation through the material (time of flight), $\Delta t$ is obtained from the transmitted waveform captured by a digital oscilloscope. In solid materials, ultrasound propagates in the form of longitudinal and shear waves. Since the shear wave travels slower and arrives later than the longitudinal wave, both longitudinal and shear wave propagation can be captured in a single waveform (Fig.~\ref{fig:waveform-time}). The signal amplitude (strength) of each waveform depends on the type of transducer used, as the transducers of specific types (longitudinal or shear) produce and respond best to their native type of waveform. Nevertheless, shear transducers will respond to longitudinal wave and vice versa.~\cite{Yurikov2019}. By selecting the right combination of transducers, it is possible to obtain both longitudinal and shear waveform from a single measurement. In this study, test sample was sandwiched between a longitudinal (Olympus,V1091) and a shear (Olympus,V157-RM) transducer with fundamental frequency 5 MHz and diameter 6.35 mm (Fig.~\ref{fig:setup-pulse transmission}). Ultrasonic waveforms were recorded using the longitudinal transducer as the transmitter and the shear transducer as the receiver. Then, another waveform was recorded by using the shear transducer as the transmitter and the longitudinal transducer as the receiver. In this way, the weak shear waveform can be identified accurately. Longitudinal and shear wave speeds were calculated using the measured respective times of flight ($\Delta t_{\rm L}$, $\Delta t_{\rm S}$), and the path length of the wave propagation through the sample (sample length). The time of flight of each longitudinal and shear waves were measured as the time gap between the highest peak position of the reference waveform (transducer-nitrile-transducer combination) and the respective peak positions of sample waveform (transducer-nitrile-sample-nitrile-transducer combination) as shown in Fig.~\ref{fig:waveform-time}.  The longitudinal, $M$ and shear, $G$ moduli were derived from the respective wave speeds and material's bulk density, $\rho_{\rm b}$ as,
\begin{equation}
\label{M-rho}
M=\rho_{\rm b} {v_{\rm L}}^2
\end{equation}
\begin{equation}
\label{G-rho}
G=\rho_{\rm b} {v_{\rm S}}^2
\end{equation}
and the bulk modulus, $K$ is related to the above moduli as
\begin{equation}
\label{K-rho}
K=M-\frac{4G}{3}.
\end{equation}

\subsection{Adsorption-ultrasonic experimental setup and procedure}
Fig.~\ref{fig:setup-adsorption-ultrasonic} shows the adsorption-ultrasonic experimental setup (inspired by Yurikov et al.~\cite{Yurikov2018} and used earlier by Ogbebor et al.~\cite{Ogbebor2023}) employed to obtain water sorption isotherms simultaneously with the ultrasonic measurements under humid air conditions. Water sorption by the sample was measured at discrete values of relative humidity, RH $= p/p_0$, the ratio of the partial vapor pressure, $p$, to the saturation vapor pressure of water, $p_0$. A series of salt solutions were used to control the RH level ranging from 12\% to 98\%. In the experimental setup, the sample-transducer assembly, a sister sample, and the salt solution were placed inside a vacuum-tight container (AnaeroPack\texttrademark{} rectangular jar, Thermo Scientific\texttrademark{} R681001) held at ambient pressure. During ultrasonic measurements, it was important to maintain a constant clamping force and sample orientation between the transducers as any variations of those could lead to errors in the measurements of signal amplitude and travel time. Therefore, the sample used for ultrasonic measurements was locked in position and the sister sample was used for the gravimetric measurements. A 40 mm box fan was used inside the container to mix the vapor-air continuously. The humidity in the container was recorded using a Vernier\texttrademark{} sensor with an accuracy of $\pm 2\%$ RH, and a resolution of 0.01\% RH. The container was placed inside a thermally insulated box (a cooler box) equipped with a finned shell-and-tube heat exchanger and a large box fan. Temperature-controlled water was pumped through the heat exchanger by a Fisherbrand\texttrademark{} Isotemp\texttrademark{} bath circulator, model 6200 R20, with the temperature set to $20.00 \pm \SI{0.03}{\degreeCelsius}$. Two cylindrical monolithic carbon xerogel samples of lengths 15.65 and 16.81 mm, and diameter 9.88 mm were used for ultrasonic and gravimetric measurements, respectively. 

Repeated measurements of ultrasonic waveforms were used to monitor the water adsorption equilibration at each RH level and the equilibration was confirmed by taking the mass measurements after a sufficiently long period ($\sim$ 24 hrs). After the RH equilibration, the target ultrasonic waveform was recorded and sister sample was weighed. Then the salt solution was replaced to adjust to the next RH value and these steps were repeated to obtain the complete sorption isotherm through adsorption and desorption routes. After that dry mass $m_{\rm dry}$ and the dry elastic properties of the sample were measured in vacuum (2 Torr). The dry mass of the 16.81 mm long sister sample was 0.9932 $\pm ~0.0002~ \rm g$.

\bibliography{manuscript}
\section{Acknowledgment}
G.Y.G., A.F.K., and A.K. thank the support from the NSF CBET-2128679 and CBET-2344923 grants. B.G. thanks the sponsors of Curtin Reservoir Geophysics Consortium.

\section{Competing interests}
The authors declare that they have no known competing financial interests or personal relationships that could have appeared to influence the work reported in this paper.

\end{document}